\documentclass[useAMS,usenatbib]{mn2e}
\usepackage{graphicx}
\usepackage{url}
\title[The role episodic accretion in low-mass star formation]
{Episodic accretion, protostellar radiative feedback, and their role in low-mass star formation}
\author[Stamatellos, Whitworth, \& Hubber]
{Dimitris Stamatellos$^{\!\  1,}$\thanks{E-mail:D.Stamatellos@astro.cf.ac.uk}, Anthony~P.~Whitworth$^{\!\  1 }$ \& David A. Hubber$^{\!\  2}$ \\
 $^1$ School of Physics \& Astronomy,Cardiff University, Cardiff, CF24 3AA, Wales, UK\\
 $^2$ Department of Physics \& Astronomy, University of Sheffield, Hounsfield Road, Sheffield S3 7RH, UK}

\begin{document}

\date{Accepted 2012 . Received 2012 October 14; in original form 2012 October 14}

\pagerange{\pageref{firstpage}--\pageref{lastpage}} \pubyear{2001-}

\maketitle

\label{firstpage}

\begin{abstract}

Protostars grow  in mass by accreting material through their discs, and this accretion is initially their main source of  luminosity. The resulting radiative feedback heats the environments of young protostars, and may thereby suppress further fragmentation and star formation. There is growing evidence that the accretion of material onto protostars is episodic rather than continuous; most of it happens in short bursts that last up to a few hundred years, whereas the intervals between these outbursts of accretion could be thousands of years. We have developed a model to include the effects of episodic accretion in simulations of star formation. Episodic accretion results in episodic radiative feedback, which heats and temporarily stabilises the disc, suppressing the growth of gravitational instabilities. However, once an outburst has been terminated,  the luminosity of the protostar is low, and the disc cools rapidly. Provided that there is enough time between successive outbursts, the disc may become gravitationally unstable and fragment. The model suggests that episodic accretion may allow disc fragmentation if  (i) the time between successive outbursts is longer than the dynamical timescale for the growth of gravitational instabilities (a few kyr), and (ii) the quiescent accretion rate onto the protostar is sufficiently low (at most a few times $10^{-7}\,{\rm M}_{\sun}\,{\rm yr}^{-1}$). We also find that after a few protostars form in the disc, their own episodic accretion events shorten the intervals between successive outbursts, and suppress further fragmentation, thus limiting the number of objects forming in the disc. We conclude that episodic accretion moderates the effect of radiative feedback from young protostars on their environments, and, under certain conditions, allows the formation of low-mass stars, brown dwarfs, and planetary-mass objects by fragmentation of protostellar discs.

\end{abstract}

\begin{keywords}
Stars: formation -- Stars: low-mass, brown dwarfs -- accretion, accretion disks -- Methods: Numerical, Radiative transfer, Hydrodynamics 
\end{keywords}

\section{Introduction}
Stars form in the dense cores of molecular clouds and grow in mass by accreting material from their surroundings. Due to the initial rotation and/or turbulence of the parent clouds discs form around protostars \citep[e.g.][]{Terebey84}. A disc grows in mass by accreting material from the envelope. The material in the disc spirals inwards and onto the protostar, provided that there is an efficient mechanism to redistribute angular momentum outwards in the disc \citep[e.g.][]{Cartwright10, Attwood09,Walch12}. 
Otherwise the disc may become sufficiently massive and cool to fragment, forming mainly  low-mass stars and brown dwarfs \citep[e.g.][]{Whitworth07,Stamatellos09}. Angular momentum can be redistributed by gravitational instabilities (GIs) or  magneto-rotational instabilities (MRI) \citep{Lin87,Balbus91,Balbus98,Lodato04}.  During this process  gravitational energy is transformed into radiation  due to viscous dissipation in the disc and at the accretion shock around the protostar. This radiation  heats the region around the protostar and may suppress subsequent fragmentation and further star formation \citep{Offner09,Bate09}.

Until recently computational studies of star formation in molecular clouds ignored the effect of radiative feedback from young protostars \citep[e.g.][]{Bate09b}. This was done merely due to computational constraints but has no physical justification, since during the initial stages after the formation of a protostar, the accretion rate onto it can be rather high, and therefore its luminosity output can also be rather high (potentially up to 100s of L$_{\sun}$, even for a $1~{\rm M}_{\sun}$ protostar).  In the last few years, more detailed computational models including the effects of radiative feedback from protostars have been developed. It has been found  \citep{Krumholz06,Bate09,Offner09,Urban10,Krumholz10, Offner10} that radiative feedback from newly formed protostars can suppress fragmentation and thus inhibit further star formation. In particular, discs around young protostars are heated and stabilised and therefore discs may not contribute significantly to the production of low-mass stars and brown dwarfs \citep{Offner09,Bate09}. \cite{Bate09} \& \cite{Krumholz11} have concluded that radiative feedback  plays a critical role in regulating the stellar initial mass function.

These studies have assumed that the accretion of material onto young protostars (and therefore their energy output) is continuous. However, there is growing observational evidence that accretion is episodic, i.e. it happens in relatively short outbursts \citep{Herbig77,Dopita78, Reipurth89, Hartmann96,Greene08}.  

FU Ori-type stars \citep{Herbig77,Greene08,Peneva10,Green11} provide the first piece of this evidence. These  stars exhibit large increase in their brightness ($\sim5-6$ mag). The increase in brightness happens within $1-10$~yr and may last for tens to a few hundred years  \citep{Hartmann96,Greene08,Peneva10}. The accretion rates during these events can be up to   $\sim 5\times 10^{-4}\,{\rm M}_{\sun}\,{\rm yr^{-1}}$; this  means that a large amount of material is delivered onto the protostar during these outbursts \cite[a few $10^{-2}\,{\rm M}_{\sun} $, e.g.][]{Bell94}.

Additional evidence for episodic accretion comes from the knots seen in many outflows from young protostars \citep[e.g.][]{Dopita78, Reipurth89}. It is believed that outflows parallel to the rotation axis of a protostar are driven by accretion of material from its disc \citep{Pudritz07,Shang07}. Therefore, periodically spaced knots in such jets are indicative of episodic accretion.

The case for episodic accretion is further supported by the observed luminosities of young protostars. These luminosities are much lower than expected if protostars accrete steadily; this is referred to as the luminosity problem \citep{Kenyon90, Evans09, Enoch09, Dunham10}. For example,  if one assumes that by the end of the Class 0 phase ($\sim10^5$ yr), half the protostar's final mass has been accumulated, then  for a final mass of $1\,{\rm M}_\odot$, the mean accretion rate onto the protostar must be $\sim5\times 10^{-6}\,{\rm M}_{\sun}\,{\rm yr}^{-1}$, and the mean accretion luminosity must be $\sim 25\,{\rm L}_{\sun}$; this is more than an order of magnitude larger than the observed average \citep[$\sim2~{\rm L_{\sun}}$; e.g.] []{Enoch09}. Part of the accretion energy  (i) may be used to power outflows and/or winds \citep{Pudritz07,Shang07,Hansen12},  (ii) may be absorbed by the protostar  \citep{Hartmann97,Commercon11}, or (iii) may be used to dissociate and ionisize the accreting gas \citep{Tan04}. However, these effects are not expected to significantly reduce the protostellar luminosity. Recent observations have revealed that  Class 0 and Class I protostars have comparable  mean luminosities \citep{Evans09,Enoch09}; this suggests that protostars accrete for longer periods than previously thought, thus the luminosity problem may be less pronounced. Nevertheless, episodic accretion is still needed to explain the low luminosities of young protostars \citep{Dunham10,Offner11,Dunham12}.

The physical cause of episodic accretion is unclear. It may be  (i) due to gravitational interactions with companions or  passing  stars that  perturb the disc and enhance accretion of material onto the protostar \citep{Bonnell92,Forgan10}, (ii) due to thermal instabilities occurring in discs \citep{Hartman85,Lin85,Bell94}, or (iii) due to gravitational instabilities \citep{Vorobyov05,Machida11}.

Another interesting possibility is that episodic accretion is caused by the complementary role of MRI and GI in discs \citep{Armitage01,Zhu07,Zhu09,Zhu09b,Zhu10, Zhu10b}. The outer disc is sufficiently cool that it is susceptible to GIs, and these result in gravitational torques that transport angular momentum outwards, thereby allowing material to spiral inwards towards the central protostar. In contrast, the inner disc  is too hot for GI, but, if it becomes hot enough, the gas is thermally ionised to a sufficient degree that it couples to the magnetic field and the MRI is activated \citep[e.g.][]{Clarke09,Rice09}; the MRI then transports angular momentum outwards, allowing matter to spiral in further and onto the protostar. As a result matter accumulates in the inner disc region, and steadily heats up until it is sufficiently ionised to activate the MRI, whereupon the accumulated matter spirals into the central protostar, giving rise to an outburst. Once most of the accumulated matter in the inner disc has been deposited onto the central protostar, the temperature of the remaining matter falls, the MRI shuts off, and the accumulation of matter resumes. 

In \cite{Stamatellos11} we incorporated the  \cite{Zhu09} model, in which MRI and GI work complementary, into radiative hydrodynamic simulations of star  formation.  We found that episodic accretion can limit the effect of radiative feedback from protostars, and therefore allow disc fragmentation, and the formation of low-mass stars and brown dwarfs. Here, we expand this work and study the effect of different model parameters; we find that episodic accretion may allow fragmentation, but the process  is self-limiting.

In Section 2, we describe the computational method used to study star formation in collapsing molecular clouds, and in Section 3 we describe in detail the phenomenological model that we have developed to include the effects of episodic accretion in hydrodynamic simulations. In Section 4, we present and compare the results of models (i) with no accretion luminosity,  (ii) with continuous accretion luminosity, and  (iii) with episodic accretion luminosity. We also examine the effect of the different parameters of the episodic accretion model. In Section 5 we discuss the conditions under which episodic accretion allows or suppresses fragmentation,  and finally in Section 6 we summarize the results of this work.

\section{Computation Method}
\label{sec:methods}

We use the SPH code {\sc seren} \citep{Hubber11b, Hubber11} to treat the gas hydro- and thermo-dynamics. The code  invokes an octal tree to compute gravity and find neighbours, multiple particle timesteps, and a Leapfrog integration scheme. The code uses time-dependent artificial viscosity \citep{Morris97}  with parameters $\alpha=1$, $\beta=2\alpha$, and a Balsara switch \citep{Balsara95}, so as to reduce artificial shear viscosity. 

The chemical and radiative processes that regulate the gas temperature are treated with the method of  \cite{Stamatellos07} \citep[see also][]{Forgan09}. The net radiative heating rate for particle $i$ is 
\begin{equation} 
\label{eq:radcool}
\left. \frac{du_i}{dt} \right|_{_{\rm RAD}} =
\frac{\, 4\,\sigma_{_{\rm SB}}\, (T_{_{\rm BGR}}^4-T_i^4)}{\bar{{\Sigma}}_i^2\,\bar{\kappa}_{_{\rm R}}(\rho_i,T_i)+{\kappa_{_{\rm P}}}^{-1}(\rho_i,T_i)}\,.
\end{equation}
The positive term on the right hand side represents heating by the background radiation field, and ensures that the gas and dust cannot cool radiatively below the background radiation temperature $T_{_{\rm BGR}}$. $\sigma_{_{\rm SB}}$ is the Stefan-Boltzmann constant, $\bar{{\Sigma}}_i$ is the mass-weighted mean column-density, and  $\bar{\kappa}_{_{\rm R}}(\rho_i,T_i)$ and ${\kappa_{_{\rm P}}}(\rho_i,T_i)$ are suitably adjusted Rosseland- and Planck-mean opacities \citep{Stamatellos07}. The method takes into account compressional heating, viscous heating, heating by the background radiation field, and radiative cooling/heating. The method has been extensively tested \citep{Stamatellos07,Stamatellos08}. Despite the fact that under certain circumstances the method gives a poor estimate of the column density through which SPH particles cool \citep{Wilkins12}, it works well within high density fragments \citep{Young:2012a}, and it reproduces the macroscopic results of 1D and 3D simulations \citep{Masunaga00,Boss79,Whitehouse06, Boley06,Cai08}.

 The gas is assumed to be a mixture of hydrogen (70\%) and helium (30\%), with  an equation of state \citep{Black75,Boley07} that accounts for the rotational and vibrational degrees of freedom of molecular hydrogen, and for the different chemical states of hydrogen and helium. The dust and gas opacity is set to  $\kappa(\rho,T)=\kappa_0\ \rho^a\ T^b\,$, where $\kappa_0$, $a$, $b$ are constants \citep{Bell94} that depend on the physical processes contributing to the opacity, for example, ice mantle melting, the sublimation of dust, molecular and H$^-$ contributions \citep[for details see][]{Stamatellos07}.

Up to densities $\rho\sim 10^{-9}\,{\rm g}\,{\rm cm}^{-3}$, the self-gravitating gas dynamics, energy equation, and associated radiation transport are treated explicitly. Wherever a gravitationally bound condensation with $\rho >10^{-9}\,{\rm g}\,{\rm cm}^{-3}$ is formed, it is presumed to be destined to form a protostar. Then, in order to avoid very small timesteps, it is replaced with a sink \citep{Bate95}, i.e.  a spherical region of radius $1\,{\rm AU}$. The sink  interacts with the rest of the computational domain only through its gravity and luminosity. Matter that subsequently flows into a sink, and is bound to it, is assimilated by the sink, and ultimately ends up in a protostar at the centre of the sink.  Normally, a sink is considered to correspond to a star forming in the simulation. In this work, we assume that a sink corresponds to a star plus a part of the star's accretion disc, i.e. the disc inside the sink radius; this part of the disc will hereafter be called the ``inner accretion disc" (IAD); it is the part of the disc that is not simulated with SPH, but is modelled phenomenologically (see Section~\ref{sec:ea.model} for details).

Radiative feedback from protostars that form during the simulation is taken into account by invoking a pseudo-background radiation field with temperature $T_{_{\rm BGR}}({\bf r})$ (see Eq.~\ref{eq:radcool}) that is a function of position relative to all the protostars present in the simulation \citep{Stamatellos07,Stamatellos09b},
 \begin{eqnarray}
T_{_{\rm BGR}}^4({\bf r})&=&\left(10\,{\rm K}\right)^4+\sum_n\left\{\frac{L_n}{16\,\pi\,\sigma_{_{\rm SB}}\,|{\bf r}-{\bf r}_n|^2}\right\}\,,
\end{eqnarray}
where $L_n$ and  ${\bf r}_n$, are the luminosity and position of the $n{^{\rm th}}$ protostar.
At all times, the luminosity of a protostar  is given by
\begin{equation}
\label{eq:starlum}
L_n=\left( \frac{M_n}{{\rm M}_{\sun}}\right)^3{\rm L}_{\sun}+f\ \frac{G M_n \dot{M}_n}{R_n}\,,
\end{equation}
where $M_n$ is the mass of the protostar, $R_n$ its radius, and $\dot{M}_n$ is the accretion rate onto it. The first term on the righthand side is the intrinsic luminosity of the protostar (due to contraction and nuclear reactions in its interior). The second term is the accretion luminosity. $f$ is the fraction of the accretion energy that is radiated away at the photosphere of the protostar, rather than being expended e.g. driving jets and/or winds \citep{Pudritz07,Shang07,Offner09}. In this work we set $f=0.75$. We also assume $R_n=3~{\rm R}_{\sun}$ is the typical radius of a young protostar \citep{Palla93}. During the initial stages of star formation the accretion luminosity dominates over the intrinsic  luminosity of the protostar. 

The above equations result in a background temperature profile around each protostar that is variable with time and dependant upon the accretion rate onto the protostar and the protostar's mass. The background temperature drops away from the protostar as $r^{-2}$, which is the expected radial temperature profile on an envelope around a young protostar.

\section{A time- dependent model of episodic accretion}
\label{sec:ea.model}

Simulations of self-gravitating hydrodynamics on the scale of molecular cloud cores (i.e. sizes from $10^4$ to $10^5\,{\rm AU}$ and masses from $0.3$ to $10\,{\rm M}_\odot$) can achieve sufficient resolution to capture the formation of discs around young protostars, and the effects of GIs in the outer regions of such discs \citep[e.g.][]{Goodwin04,Bate09,Bate12,Attwood09}. However, such simulations cannot capture what happens in the inner disc region (within a few AU from the central star). In this region numerical (artificial) viscosity is larger than the viscosity due to the effect of GIs \citep[e.g.][]{Clarke09,Zhu09,Cartwright10,Forgan:2011a}, and hence matter is driven towards the central protostar at an artificially enhanced rate. Additionally, to avoid short timesteps and computationally prohibitive running times, sinks on the order of 1~AU are invoked.  It is  normally assumed that any matter that enters the  sink flows instantaneously onto the central protostar. However, at this scale magnetic fields could be important in determining the rate at which matter is delivered to the protostar, and their effect needs to be included in the models.
 
To take into account the physics that is dominant in the inner disc region, without increasing greatly the computational cost, we have incorporated the results of established semi-analytic models \citep{Zhu09,Zhu09b,Zhu10,Zhu10b} into hydrodynamic simulations of star formation \citep{Stamatellos11}. We assume that  the mass accreted into a sink  is deposited initially onto an inner accretion disc  (IAD) inside the sink, where it piles up until it becomes hot enough that thermal ionisation couples the matter to the magnetic field. At this point the MRI is activated, transporting angular momentum outwards, and thereby allowing the matter accumulated in the IAD to spiral inwards and onto the central protostar. Hence the sink mass is divided between the central protostar ($M_\star$) and its IAD ($M_{_{\rm IAD}}$):
\begin{equation}
M_{_{\rm SINK}}=M_{\star}+M_{_{\rm IAD}}\,.
\end{equation}
Matter is assumed to flow from the IAD onto the protostar at a rate
\begin{equation}
\dot{M}_{\star}=\dot{M}_{_{\rm BRG}}+\dot{M}_{_{\rm MRI}}\,,
\end{equation}
where $\dot{M}_{_{\rm BRG}}$ is the quiescent accretion rate, i.e. when there is no outburst, and the second term is a much higher accretion rate that obtains only when MRI acts to transport angular momentum in the IAD. { The IAD is presumed to be a continuation of the much more extended accretion disc outside the sink, which is simulated explicitly.}

In detailed disc models \citep{Zhu09,Zhu09b,Zhu10,Zhu10b} the matter in the IAD couples to the magnetic field, due to thermal ionisation, once the temperature reaches a threshold of $T_{_{\rm MRI}}\sim 1400\,{\rm K}$.  Using an $\alpha$-parameterization \citep{Shakura73} for the effective viscosity delivered by the MRI, \cite{Zhu10} estimate that the accretion rate during an outburst is
\begin{equation}
\label{eq:mrimdot}
\dot{M}_{_{\rm MRI}}\sim 5\times10^{-4}\,{\rm M}_{\sun}\,{\rm yr}^{-1}\,\left(\frac{\alpha_{_{\rm MRI}}}{0.1}\right)\,,
\end{equation}
where $\alpha_{_{\rm MRI}}$ is the effective Shakura-Sunyayev  parameter \citep{Shakura73} for the  viscosity provided by the MRI.  \cite{Zhu10} also find that the duration of an outburst is
\begin{eqnarray}
\label{eq:mridt}
\Delta t_{_{\rm MRI}} &\,\sim\,0.25\,{\rm kyr}&\left(\frac{\alpha_{_{\rm MRI}}}{0.1}\right)^{-1}\,
\left(\frac{M_{\star}}{0.2{\rm M}_{\sun}}\right)^{2/3}\times\nonumber\\ 
& &
\left(\frac{\dot{M}_{_{\rm IAD}}}{10^{-5}\ {\rm M}_{\sun}\,{\rm yr}^{-1}}\right)^{1/9}\,.
\end{eqnarray}
Here $\dot{M}_{_{\rm IAD}}$ is the rate at which matter flows into the sink and onto the IAD.
We assume that the critical temperature  for the MRI to be activated is reached when enough mass for an MRI-enabled outburst has been accumulated in the IAD, i.e. 
\begin{equation}
\label{eq:m.mri}
M_{_{\rm IAD}}\;\,>\;\
M_{_{\rm MRI}}
\;\,\sim\;\,{\dot{M}_{_{\rm MRI}}}{\Delta t_{_{\rm MRI}}}\,.
\end{equation}
Using Eqs.~\ref{eq:mrimdot} and \ref{eq:mridt} we obtain
\begin{equation}
M_{_{\rm IAD}}\!>\! 0.13\,{\rm M}_{\sun}\!
\left(\frac{M_{\star}}{0.2{\rm M}_{\sun}}\right)^{2/3}\!
\left(\frac{\dot{M}_{_{\rm IAD}}}{10^{-5}\,{\rm M}_{\sun}\,{\rm yr}^{-1}}\right)^{1/9}.
\end{equation}

The detailed  thermal and ionisation balance in the IAD is not modeled explicitly since this is not the main concern of this paper. Based on observations and models of FU Ori-type stars \citep{Hartmann96,Zhu10}, we assume that rapid accretion onto the protostar, at a rate given by
\begin{eqnarray}
\label{eq:dmdtdrop}
\dot{M}_{\rm MRI}\!=1.58\, \!\frac{M_{_{\rm MRI}}}{\Delta t_{_{\rm MRI}}}\!\exp\!\left\{\!-\!\frac{(t\!-\!t_0)}{\Delta t_{_{\rm MRI}}}\right\},t_{_{\rm O}}\!<\!t\!<\!t_{_{\rm O}}\!+\!\Delta t_{_{\rm MRI}}
\end{eqnarray}
is initiated as soon as $M_{_{\rm IAD}}$ exceeds $M_{_{\rm MRI}}$; $t_{_{\rm O}}$ is the time at which this occurs, and the outburst terminates at $t_{_{\rm O}}+\Delta t_{_{\rm MRI}}$. The factor $1.58=e/(e-1)$ is to allow for all the mass of the IAD  to be accreted onto the star within $\Delta t_{_{\rm MRI}}$.

The time taken to accumulate (or re-accumulate) $M_{_{\rm MRI}}$ is $\Delta t_{_{\rm ACC}} \sim M_{_{\rm MRI}}/\dot{M}_{_{\rm IAD}}$, i.e. 
\begin{equation}\label{eq:dtACC}
\Delta t_{_{\rm ACC}}\!\simeq\!13{\rm kyr}\!\left(\!\frac{M_\star}{0.2{\rm M}_\odot}\!\right)^{2/3}\!\left(\!\frac{\dot{M}_{_{\rm IAD}}}{10^{-5}{\rm M}_\odot\,{\rm yr}^{-1}}\!\right)^{-8/9}\,.
\end{equation}
This is effectively the interval between successive outbursts, and, as will be explained later in the paper, crucially  it is much longer than the duration of the outburst (see Eq.~\ref{eq:mridt}), and significantly longer than  the growth time of GIs  (a few kyr).

In this formulation there are two free parameters. The first is $\alpha_{_{\rm MRI}}$, which controls the strength and duration of the outburst; increasing $\alpha_{_{\rm MRI}}$ makes for a more intense, shorter outburst. The mass delivered in each outburst, and the duration of the interval between outbursts, are essentially independent of  $\alpha_{_{\rm MRI}}$, which is fortunate, in the sense that $\alpha_{_{\rm MRI}}$ is rather uncertain. Observations and  simulations suggest  $\alpha_{_{\rm MRI}}\!=\!0.01\;{\rm to}\;0.4$ \citep[e.g.][]{King07,Isella:2009a}. The second free parameter is the quiescent accretion rate onto the protostar, $\dot{M}_{_{\rm BRG}}$, i.e. the rate at which material is assumed to accrete onto the protostar when the MRI is not active.

\begin{table*}
\begin{minipage}{\textwidth}
\caption{Summary  of seven simulations of collapsing turbulent cores; with no radiative feedback from protostars (run ea0), with continuous radiative feedback (run ea1), and with episodic radiative feedback with various parameters (runs ea2-ea6).  $\alpha_{\rm MRI}$ is the viscosity parameter due to MRI, 
$\dot{M}_{_{\rm BRG}}$ is the quiescent accretion rate onto the protostar, $N_f$ is the number of objects (stars and brown dwarfs) formed in the simulation, $M_i$ their masses (at the end of the simulations, i.e. at 100~kyr), and $t_i$ their formation time. For the simulations with episodic accretion, and hence episodic radiative feedback, we also list the percentage of a star's life during which it is experiencing high MRI-driven accretion ($t_{\rm EA,i }/t_{\star,i}$). Multiple entries of $M_i$, $t_i$, and  $t_{\rm EA,i }/t_{\star,i}$ refer to different objects formed, in order of their formation time.}
\label{tab:summary}
\centering
\renewcommand{\footnoterule}{}  
\begin{tabular}{@{}cccclllc} \hline
\noalign{\smallskip}
run id 	&	$\alpha_{_{\rm MRI}}$ & $\dot{M}_{_{\rm BRG}}$ (${\rm M}_{\sun}\,{\rm yr}^{-1}$)
&  $N_f$ & $M_i$ (M$_{\odot}$) & $t_i$ (kyr) & $t_{\rm EA, i }/t_{\star,i}$ (\%) \\
\noalign{\smallskip}
\hline
\noalign{\smallskip}
ea0  		 & -  	 &- 			& 4 & 0.49, 0.32, 0.005, 0.23 & 78.7, 87.2, 88.8, 90.0 & -\\
ea1  		&  -	 & -			& 1 & 0.59 & 78.7 & \\
ea2  	 	& 0.1 & $10^{-7}$	& 2 & 0.63, 0.37&78.7, 89.9 &4, 6 \\
ea3   		& 0.01& $10^{-7}$	& 1 & 0.60 & 78.7 & 41  \\
ea4   	 	& 0.3  & $10^{-7}$	& 3 & 0.63, 0.26, 0.12&78.7, 90.7, 91.0 &1, 2, 1 \\
ea5  		& 0.1  & $10^{-8}$	& 3 & 0.56, 0.35, 0.11 &78.7, 89.0, 90.9  &5, 2 \\
ea6 		& 0.1  & $10^{-6}$	& 1 & 0.64 & 78.7 & 4\\

 \noalign{\smallskip}
\hline
\end{tabular}
\end{minipage}
\end{table*}

\section{The role of episodic accretion in low-mass star formation}

To determine the role of episodic accretion in star formation, we simulate the collapse of a star forming cloud core and the subsequent genesis of protostars. The initial properties of the molecular cloud core \citep{Goodwin04,Goodwin04b,Goodwin06, Attwood09}  are chosen to match  the observed properties of prestellar cores \citep{WardThompson94, Andre96,WardThompson99,Andre00, Alves01, Kirk05, Stamatellos03,Stamatellos04}. The initial density profile is
\begin{eqnarray}
\rho(r)&=&\frac{\rho_{_{\rm KERNEL}}}{\left(1\,+\,\left(r/R_{_{\rm KERNEL}}\right)^2\right)^2}\,.
\end{eqnarray}
Here $\rho_{_{\rm KERNEL}}=3\times 10^{-18}\,{\rm g}\,{\rm cm}^{-3}$ is the central density, and $R_{_{\rm KERNEL}}=5,000\,{\rm AU}$ is the radius of the central region within which the density is approximately uniform; the free-fall time at the centre is $\sim40$~kyr.
The outer envelope of the core extends to $R_{_{\rm CORE}}=50,000\,{\rm AU}$, so the total core mass is $M_{_{\rm CORE}}=5.4\,{\rm M}_\odot$, and the mass initially inside $R_{_{\rm KERNEL}}$ is $M_{_{\rm KERNEL}}=1.1\,{\rm M}_\odot$. The gas is initially isothermal at $T=10\,{\rm K}$, and hence the initial ratio of thermal to gravitational energy is $\alpha_{_{\rm THERM}}\equiv{U_{_{\rm THERM}}}/{|U_{_{\rm GRAV}}|}=0.3\,.$ We impose an initial random, divergence-free, turbulent velocity field \citep{Bate09}, with power spectrum $P_kdk\propto k^{-4}dk$, to match the scaling laws observed in molecular clouds \citep{Larson81,Burkert00}, and amplitude such that $\alpha_{_{\rm TURB}}\equiv{U_{_{\rm TURB}}}/{|U_{_{\rm GRAV}}|}=0.3\,$\citep{Jijina99}.

The molecular cloud core is represented by $10^{6}$ SPH particles, so each SPH particle has mass $m_{_{\rm SPH}}\simeq 5\times 10^{-6}\,{\rm M}_\odot$. The minimum resolvable mass is therefore $M_{_{\rm MIN}}\simeq {\cal N}_{_{\rm NEIB}}m_{_{\rm SPH}}\simeq 3\times 10^{-4}\,{\rm M}_\odot$. This is much smaller than the theoretical minimum mass for star formation \citep[$\sim 3\times 10^{-3}\,{\rm M}_\odot$, usually referred to as {\it the opacity limit}, e.g.][]{Whitworth06}, and therefore all self-gravitating condensations are well resolved.

\begin{figure*}
\centerline{
\includegraphics[height=\textwidth,angle=-90]{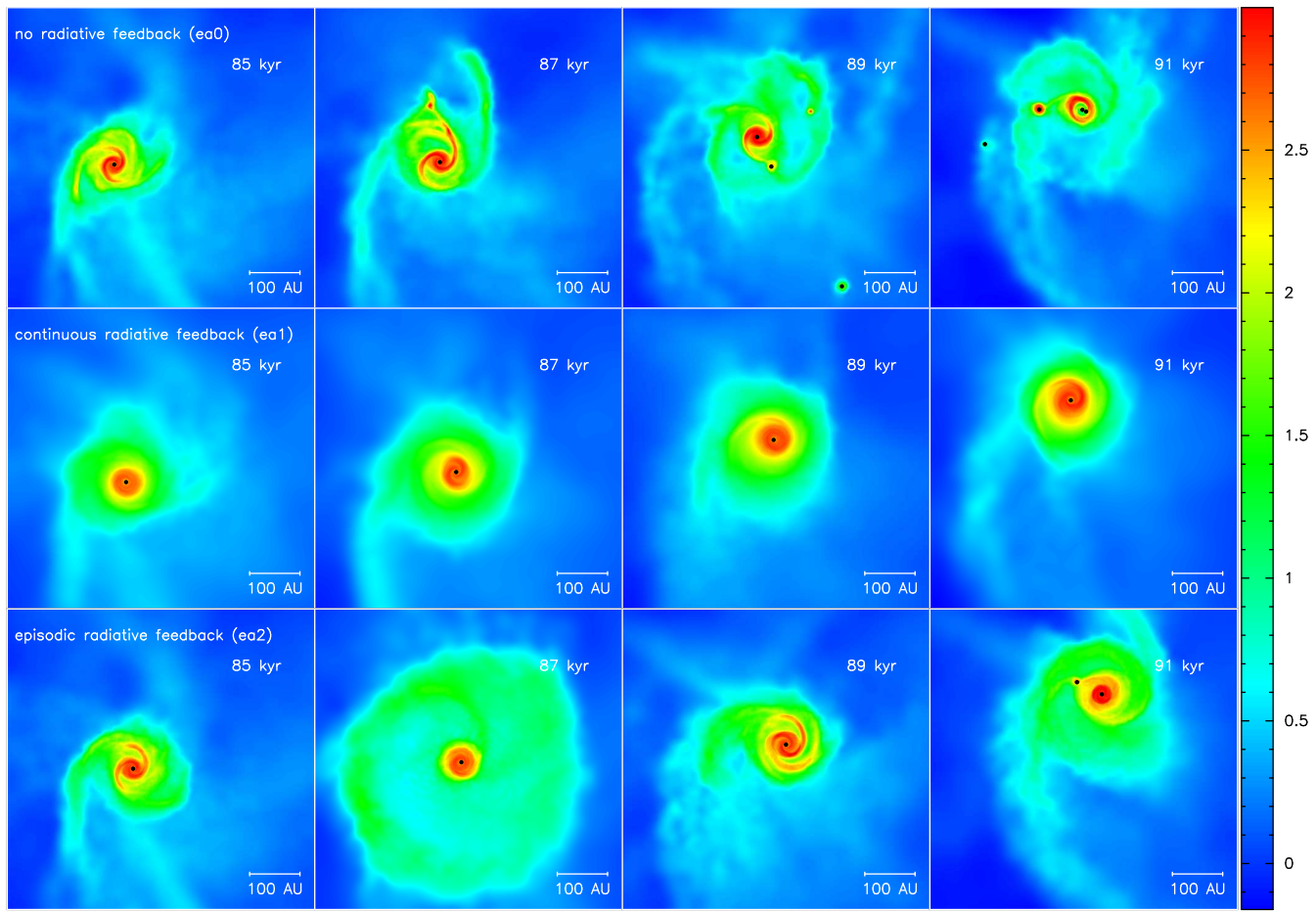}}
\caption{Evolution of the accretion disc around the primary protostar forming in a collapsing turbulent molecular cloud core. The colour encodes the logarithm of column density, in ${\rm g\ cm}^{-2}$. The different rows present time sequences from 85 to $91\,{\rm kyr}\,$ for different treatments of the protostellar radiative feedback. Top row (ea0), no radiative feedback: the disc around the primary protostar increases in mass, becomes gravitationally unstable, and fragments to form 2 low-mass stars and 1 planetary-mass object. Middle row (ea1), continuous accretion and continuous radiative feedback: the disc grows in mass, but radiative feedback keeps it sufficiently hot so that it does not fragment. Bottom row (ea2), episodic accretion and episodic radiative feedback: the disc becomes gravitationally unstable (first  column) but the growth of GIs is suppressed by heating due to an accretion burst (second column); however, after this burst the disc cools again, GIs quickly develop, and eventually the disc fragments; one low-mass star forms in the disc.}
\label{fig:snapshots}
\end{figure*}

\begin{figure}
\centerline{
\includegraphics[height=\columnwidth,angle=-90]{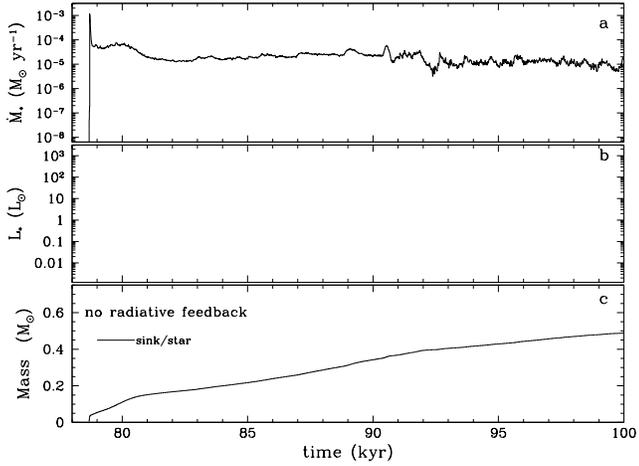}}
\centerline{
\includegraphics[height=\columnwidth,angle=-90]{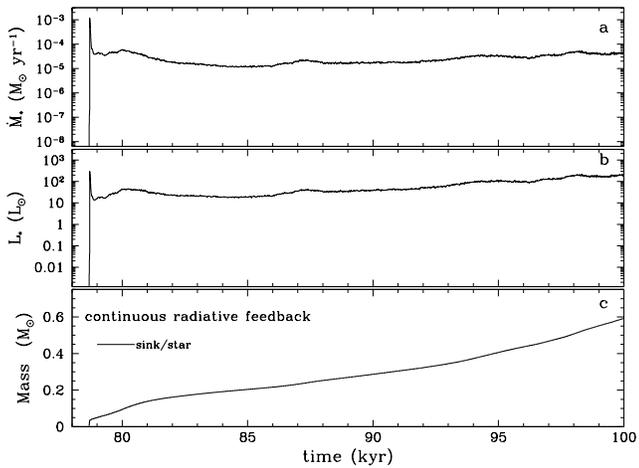}}
\centerline{
\includegraphics[height=\columnwidth,angle=-90]{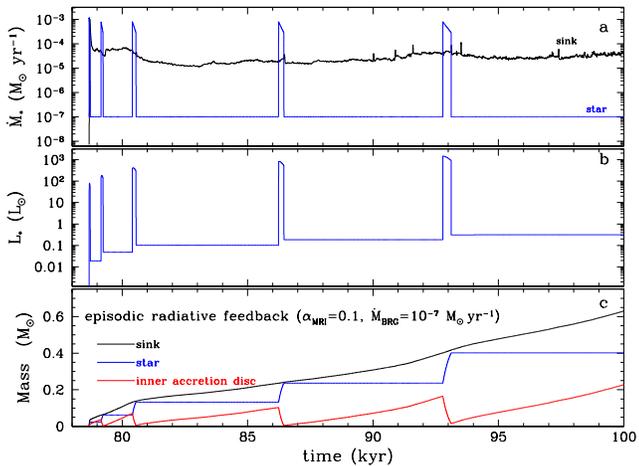}}
\caption{The effect of radiative feedback: (a) the accretion rates, (b) the luminosities, and (c) the sink/star/IAD masses of the primary protostar formed in the simulation, (i) with no radiative feedback (run ea0; top panel), (ii) with continuous radiative feedback (run  ea1; middle panel), and (iii) with episodic radiative feedback (run ea2; bottom panel), with {$\alpha_{_{\rm MRI}}\!=\!0.1$, $\dot{M}_{_{\rm BRG}}=10^{-7}\,{\rm M}_{\sun}\,{\rm yr}^{-1}$}. In the first two simulations 'sink' and 'star' are equivalent, whereas in the episodic accretion run a sink comprises the star and its inner accretion disc. In the run with continuous radiative feedback the accretion luminosity stays high at $\sim 10-200~{\rm L}_{\sun}$, whereas in the run with episodic radiative feedback due to episodic accretion of material onto the protostar, the luminosity stays low ($\stackrel{_<}{_\sim} 1~{\rm L}_{\sun}$), apart from during the episodic events in which the luminosity rises up to $\sim10^3~{\rm L}_{\sun}$.}
\label{fig:group.a}
\end{figure}

\begin{figure}
\centerline{
\includegraphics[height=\columnwidth,angle=-90]{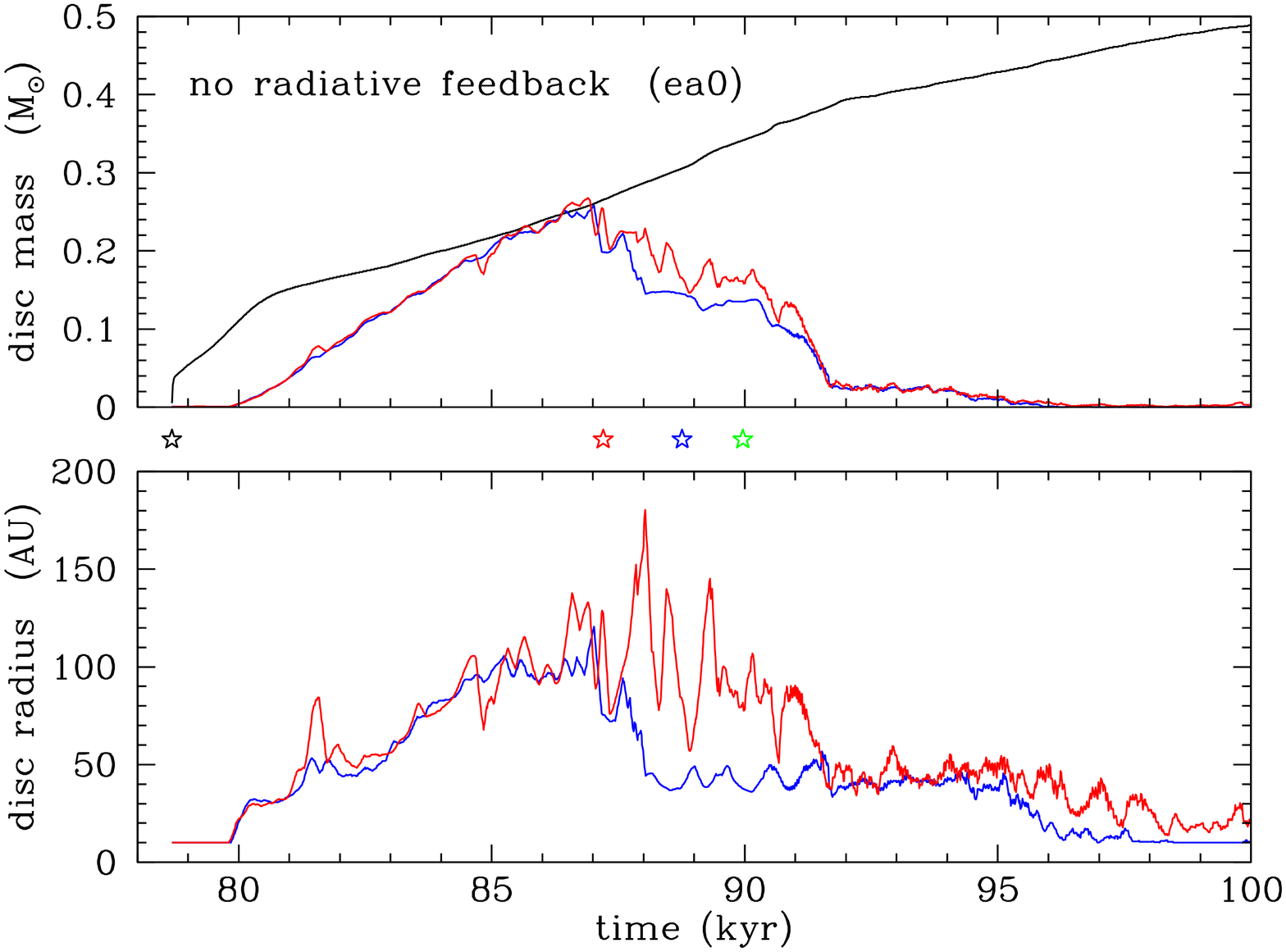}}
\centerline{
\includegraphics[height=\columnwidth,angle=-90]{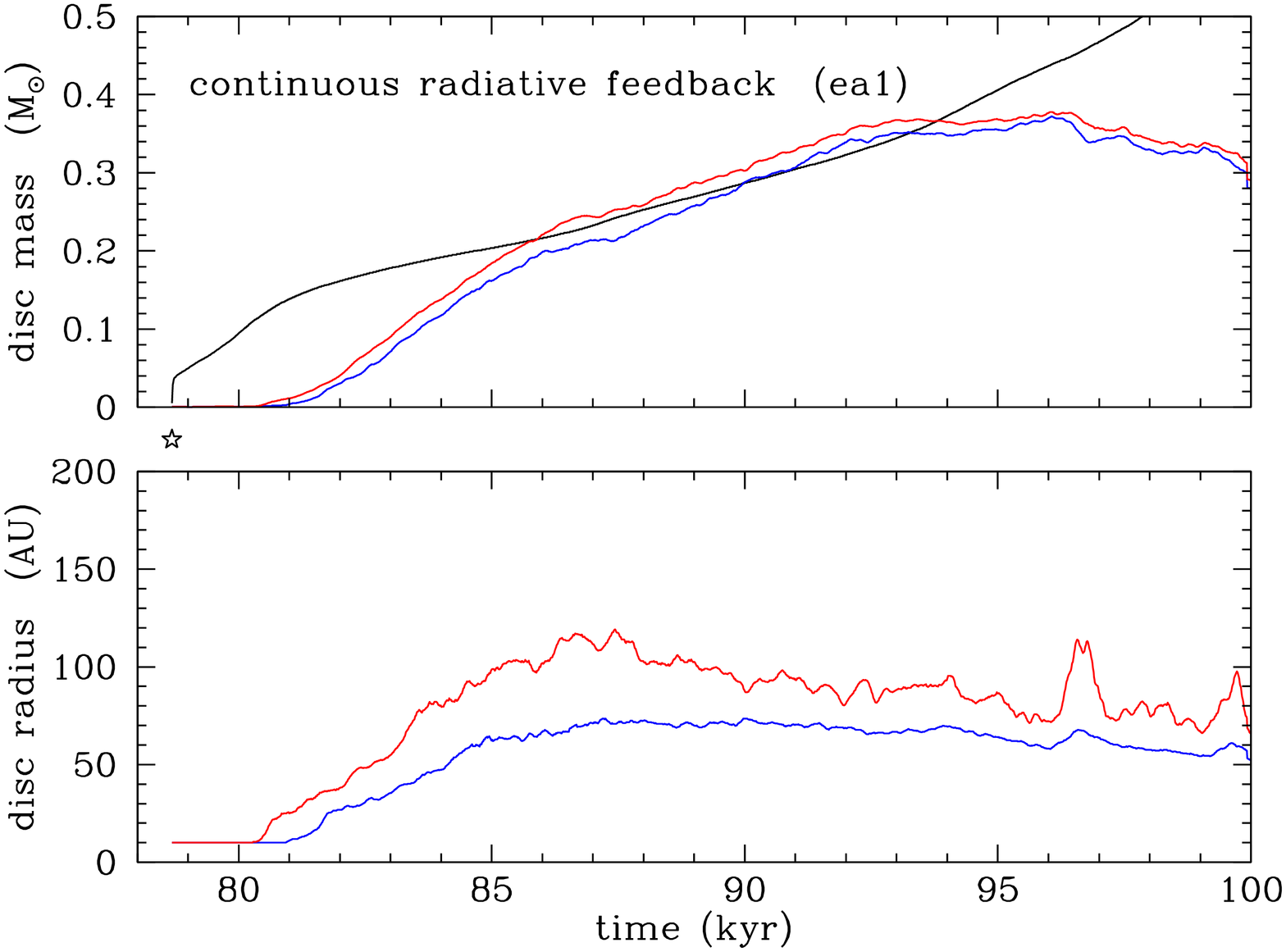}}
\centerline{
\includegraphics[height=\columnwidth,angle=-90]{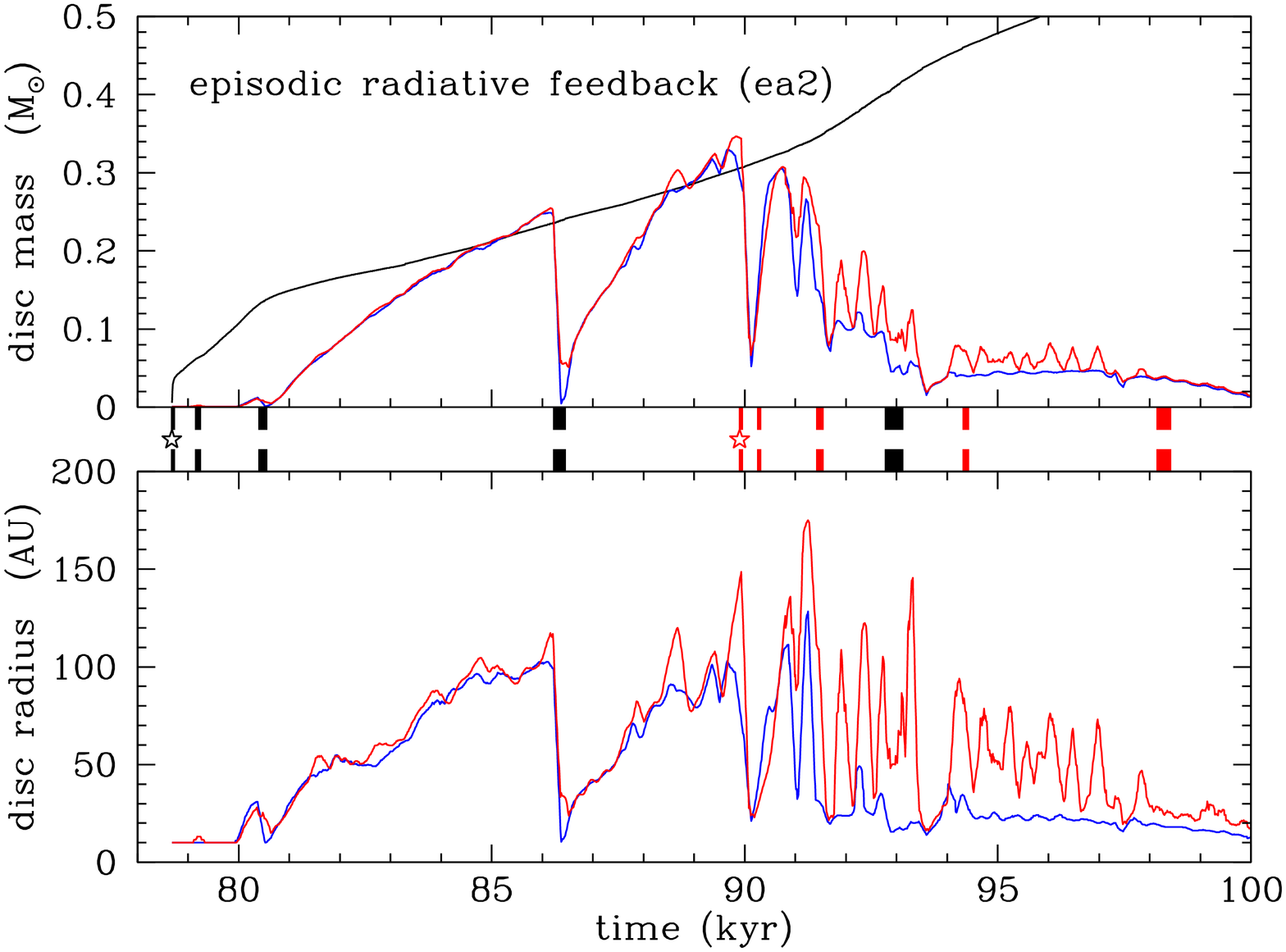}}
\caption{The estimated disc  masses and radii calculated using (a)  $v_r(r_i)<2\pi v_\phi(r_i)$ (red lines), and
(b) $v_\phi(r_i)>0.9\ v_k(r_i)$  (blue lines),  for the run (i) with no radiative feedback (ea0; top panel), (ii) with continuous radiative feedback (ea1; middle panel), and (iii) with episodic radiative feedback (ea2; bottom panel). The black lines denote the total mass of the (primary) sink in each simulation (for runs ea0 and ea1 this corresponds to the primary protostar; for run ea2 this corresponds to the protostar plus the IAD). Star signs in between graphs mark the formation time of primary and secondary protostars. In the run with episodic accretion (ea2) the bands in between graphs mark the intervals where episodic accretion occurs (black: episodic accretion onto the primary star, red: episodic accretion onto a secondary protostar). The estimates of disc masses and radii are reliable only until secondary stars form in the discs. Episodic accretion events disrupt the disc structure (bottom panel; $t\sim86$~kyr). }
\label{fig:discs}
\end{figure}

We perform seven simulations, all with the same initial conditions, and differing only in their treatments of the luminosities of protostars (see Table~\ref{tab:summary}, Figs.~\ref{fig:snapshots}-\ref{fig:toomre.c}). In the first simulation  (ea0) there is no radiative feedback from young protostars, and in the second simulation (ea1) the protostellar radiative feedback is continuous. In the remaining  simulations (ea2-ea6) episodic accretion is invoked, which leads to episodic radiative feedback. Each run uses different values for the MRI viscosity parameter, $\alpha_{_{\rm MRI}}$, and/or the quiescent accretion rate onto the protostar, $\dot{M}_{_{\rm BRG}}$.  

In all simulations the initial collapse leads to the formation of a protostar at $\,t\!\sim\!79\,{\rm kyr}$, and this  acquires an accretion disc.  The simulations only diverge after this juncture, as the disc  growth and evolution depend on how the protostellar feedback is treated. The simulations are run until 100~kyr.

It is not trivial to define a disc forming in a hydrodynamical simulation and to calculate its properties \citep{Machida10,Walch12}. Firstly, we calculate the angular momentum vector of the sink/protostar, and assume that the disc plane is perpendicular to it. Then we define concentric rings around the sink,  we assume that the  disc has thickness of $z=0.2r$ (note that since there is little mass outside  $0.1r$ from the disc midplane the estimated disc properties are not significantly affected by the choice of the disc thickness), and calculate the (azimuthal) average of the radial ($v_r$) and azimuthal ($v_\phi$) velocity. We also estimate the Keplerian velocity at the radius of each ring, using $v_{k}(r_i)=(GM_i/r_i)^{1/2}$, where $r_i$ is the radius of the ring, and $M_i$ the mass inside this radius (including the mass of the sink). We calculate the disc radius (and afterwards the disc mass) in two ways: (i) we estimate the most distant ring for which $v_r(r_i)<2\pi v_\phi(r_i)$, i.e.  matter can go around the sink at least once before it falls onto the sink (Fig.~\ref{fig:discs}, red lines), and (ii) we estimate the most distant ring for which $v_\phi(r_i)>0.9\ v_k(r_i)$, i.e. the rotation velocity is nearly Keplerian (Fig.~\ref{fig:discs}, blue lines). Both criteria  give comparable results for the disc mass and radius,  although the disc is generally somewhat more compact with the latter criterion (see Fig.~\ref{fig:discs}).

The details of the simulations and their results in terms of star formation are summarised in Table~\ref{tab:summary}. Henceforth, we will refer to the first object formed in a simulation as the {\it primary protostar}, whereas the rest, which are formed in the disc of the primary protostar,  for simplicity will be referred to collectively as {\it secondary protostars} (even though some of them have masses below the hydrogen burning limit, $\sim80$~M$_{\rm J}$,  and therefore are brown dwarfs and/or planetary-mass objects).

\subsection{No radiative feedback}
In this run (ea0) mass accretes continuously onto the protostar (see Figs.~\ref{fig:snapshots}, \ref{fig:group.a}; top panels), but this does not translate into  radiative feedback, as  the luminosity of the protostar in Eq.~\ref{eq:starlum} is artificially set to 0.  The disc around the primary protostar grows in mass (up to $\sim 0.2~{\rm  M}_{\sun}$; see Fig.~\ref{fig:discs}, top) and becomes larger (radius out to $\sim 100$~AU)  \citep[cf.][]{ Zhu12}. 
It becomes gravitationally  unstable and fragments at radii $>50$~AU, where it can cool fast enough \citep{Matzner05,Whitworth06,Whitworth07,Stamatellos07b, Kratter08,Stamatellos09,Stamatellos09b}. The temperature at a distance of 100 AU from the primary protostar, a characteristic radius at which fragmentation is expected, remains very low around 10~K (Fig.~\ref{fig:toomre.a}a, red line).  The Toomre parameter $Q$ at the same radius drops to near unity as the disc grows in mass, indicating that the disc becomes gravitationally unstable (Fig.~\ref{fig:toomre.a}b, red line). Three secondary objects form in the disc (see Table~\ref{tab:summary}); two of them are low-mass stars and the third is a planetary-mass object. The planetary-mass object is quickly ejected from the disc, as seen in previous simulations \citep{Stamatellos09}; this is critical for terminating  its growth by accreting material from the disc \citep{Stamatellos09b}. The results presented here are consistent with previous studies that, in the absence of protostellar radiative feedback, report disc fragmentation and low-mass star formation \citep{Bate02, Goodwin04,Attwood09,Bate09b,Offner09}.

\subsection{Continuous radiative feedback}
In this run (ea1; Fig.~\ref{fig:snapshots}, middle panel) mass accretes continuously onto the protostar, and gravitational energy is radiated away according to Eq.~\ref{eq:starlum}. Therefore the protostar continuously heats its environment.  The luminosity of the protostar varies from $10-200~{\rm L}_{\sun}$ (Fig.~\ref{fig:group.a}, middle panel). As in the previous case the disc around the primary protostar grows in mass (up to $\sim 0.4~{\rm  M}_{\sun}$; see Fig.~\ref{fig:discs}, middle) and extends in radius (out to $\sim 100$~AU). The disc is  continuously heated by the protostar and therefore it is stabilised against fragmentation, despite its large mass. The temperature at 100~AU remains relatively high at around 50 K at all times (Fig.~\ref{fig:toomre.a}a, blue line), and the corresponding Toomre parameter is always larger than 2 (Fig.~\ref{fig:toomre.a}b, blue line), indicating a stable disc. There are occasions when the disc becomes marginally unstable and spiral arms start developing in the disc; this leads to more material driven inwards in the disc and eventually onto the protostar, and therefore more energy is released that heats and stabilises the disc, dampening any spiral structure. The disc is therefore in a self-regulating stable state \citep{Lodato04,Lodato05}.

These results agree with the studies of  \cite{Bate09} \& \cite{Offner09}, which show that when protostellar radiative feedback is included, disc fragmentation is generally suppressed and  fewer low-mass stars and brown dwarfs form.

\subsection{Episodic radiative feedback}

In this run (ea2; {$\alpha_{_{\rm MRI}}\!=\!0.1$, $\dot{M}_{_{\rm BRG}}=10^{-7}\,{\rm M}_{\sun}\,{\rm yr}^{-1}$}; Fig.~\ref{fig:snapshots}, bottom panel) mass accretes onto the protostar episodically using the model described in detail in Section~\ref{sec:ea.model}.  The mass flows into the inner disc region and then it accretes onto the protostar only when the conditions are right for the MRI to operate. The accretion rate is low ($10^{-7}\,{\rm M}_{\sun}\,{\rm yr}^{-1}$) when the MRI is inactive, and high  ($\sim10^{-3}\,{\rm M}_{\sun}\,{\rm yr}^{-1}$) when the MRI is active. Therefore, the release of energy from the protostar is also episodic; for most of the time it is below $1~{\rm L}_{\sun}$, but during episodic accretion events it can get up to $\sim10^3~{\rm L}_{\sun}$ (Fig.~\ref{fig:group.a}, bottom panel). However, the protostellar luminosity does not suppress fragmentation, since it it high only for relatively short periods of time. 

The disc grows in mass and radius as in the case with no radiative feedback (mass up to $\sim 0.2~{\rm  M}_{\sun}$  and  radius out to $\sim 100$~AU; see Fig.~\ref{fig:discs}, bottom), and the disc becomes marginally unstable. At this instant ($t\sim86$~kyr) an episodic accretion event occurs (see Fig.~\ref{fig:snapshots}, bottom panel, second snapshot) disrupts the disc structure, and seems to suppress the growth of gravitational instabilities. However, this is only temporary; after the episodic accretion event terminates, the disc quickly becomes gravitationally unstable again. This time the disc remains cool for long enough to fragment and produce a low-mass H-burning star (see Fig.~\ref{fig:snapshots}, bottom panel, last two snapshots). 

\begin{figure}
\centerline{
\includegraphics[height=\columnwidth,angle=-90]{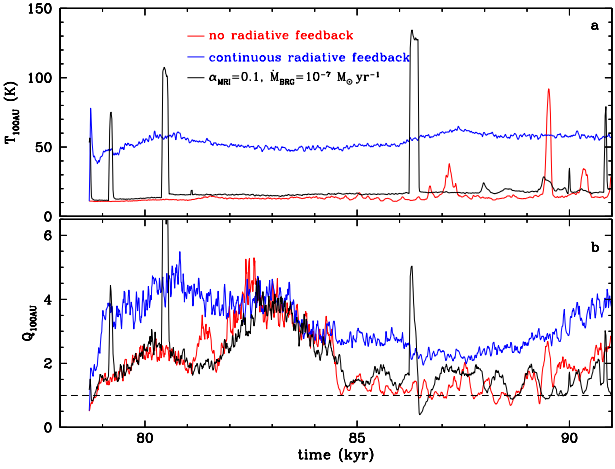}}
\caption{The azimuthally-averaged (a) temperature (top), and (b) Toomre parameter $Q$ (bottom) at radius 100~AU from the primary protostar plotted against time, for the run with no radiative feedback (ea0; red), with continuous radiative feedback (ea1; blue), and with episodic radiative feedback (ea2; black). }
\label{fig:toomre.a}
\end{figure}

At the characteristic radius of 100 AU the azimuthally averaged temperature  (Fig.~\ref{fig:toomre.a}a, black line) is very similar to the corresponding temperature of the run with no radiative feedback ($\sim10$~K), except during episodic accretion events, when the temperature may go up to more than 100~K. Once the outburst terminates the disc quickly cools down to $\sim~10$~K. Assuming that the disc surface density at 100~AU is $\Sigma\sim20\ {\rm g\ cm}^{-2}$,  the temparature $T\sim100$~K, and the corresponding Rosseland-mean opacity $\kappa\sim1\  {\rm cm^{2} g^{-1}}$, using  Eq.~\ref{eq:radcool} we find that the cooling rate is $\dot{u}_{\rm c}\sim50\ {\rm erg\ s^{-1}g^{-1}}$. Additionally, assuming that the disc mass  is $M_{\rm disc}\sim0.1\ {\rm M}_{\sun}$ we find that the total specific internal energy of the disc is $u=(3/2)(kT/\mu m_{_{\rm H}})\sim6\times10^9\ {\rm erg\ g^{-1}}$. Therefore, the timescale of disc cooling is $t_{\rm c}=u/\dot{u}_{\rm c}\sim 4$~yr, i.e. much shorter than the interval between outbursts.

The Toomre parameter $Q$ is also similar to the case with no radiative feedback except during the episodic accretion events (Fig.~\ref{fig:toomre.a}b, black line). The interval between successive episodic accretion events is on the order of  a few thousand years ($\sim 6$~kyr), which is adequate time for gravitational instabilities to grow and the disc to fragment.  

\subsection{The effect of different parameters for the episodic accretion model}

The episodic accretion model (Section~\ref{sec:ea.model}) has 2 free parameters;  (i) the viscosity parameter $\alpha_{_{\rm MRI}}$, which controls the strength and duration of the outburst;  (ii)  the quiescent accretion rate $\dot{M}_{_{\rm BRG}}$, i.e. the rate at which material is assumed to accrete onto the protostar when the MRI is not active. These are rather uncertain. Observations and  simulations of discs suggest  $\alpha_{_{\rm MRI}}\!=\!0.01\;{\rm to}\;0.4$ \citep[e.g.][]{Balbus98,King07,Isella:2009a}, whereas observations of young stars suggest quiescent accretion rates around $10^{-11}-10^{-6}\,{\rm M}_{\sun}\,{\rm yr}^{-1}$ \citep[e.g.][]{Calvet04,Natta04, Muzerolle05,Mohanty05}. In the following sections we examine the effect of these parameters on the growth of gravitational instabilities and disc fragmentation.

\subsubsection {\bf $\alpha_{_{\rm MRI}}\!=\!0.01,0.1,0.3$  ($\dot{M}_{_{\rm BRG}}=10^{-7}\,{\rm M}_{\sun}\,{\rm yr}^{-1})$}

\begin{figure}
\centerline{
\includegraphics[height=\columnwidth,angle=-90]{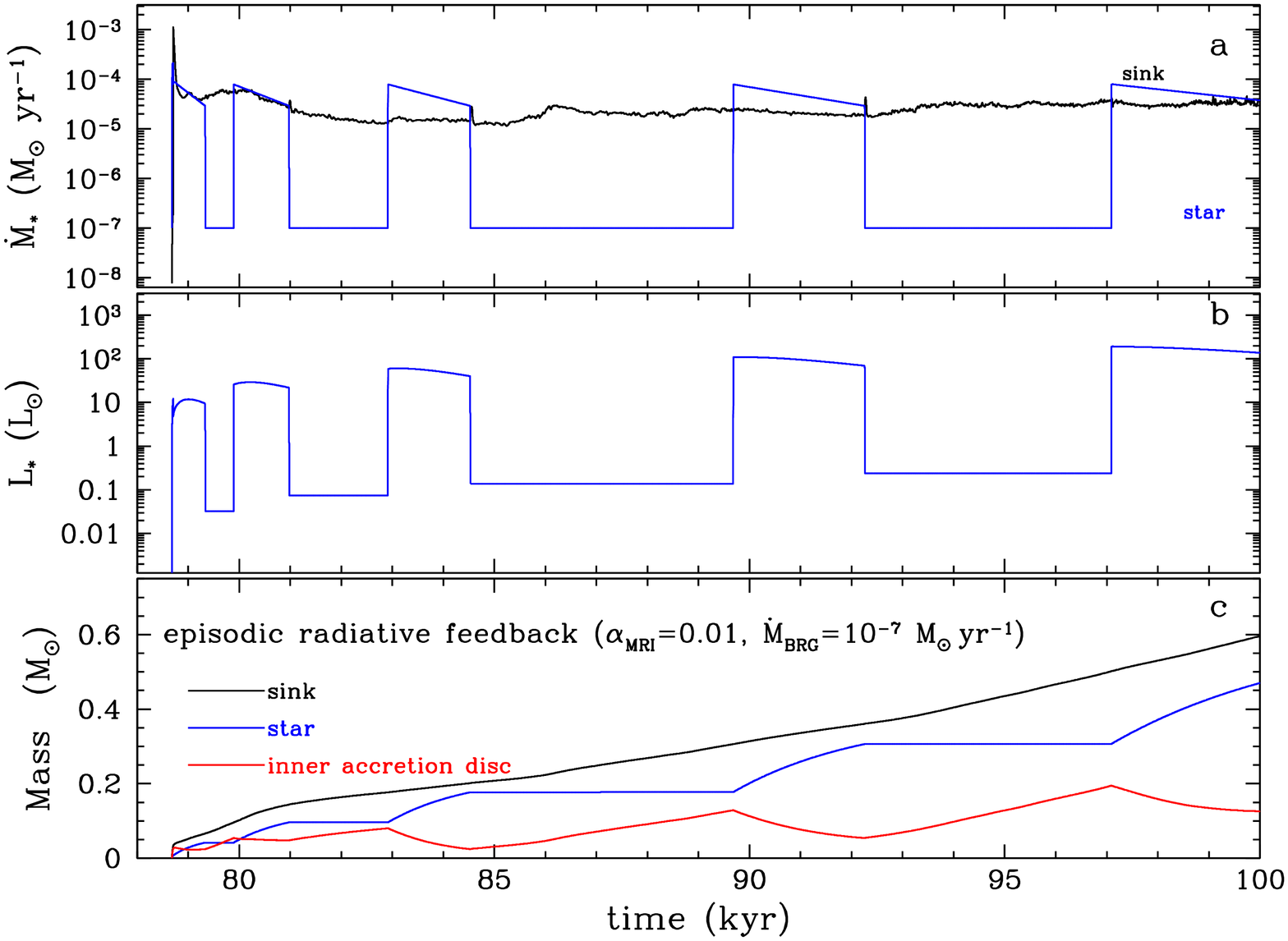}}
\centerline{
\includegraphics[height=\columnwidth,angle=-90]{ea2.eps}}
\centerline{
\includegraphics[height=\columnwidth,angle=-90]{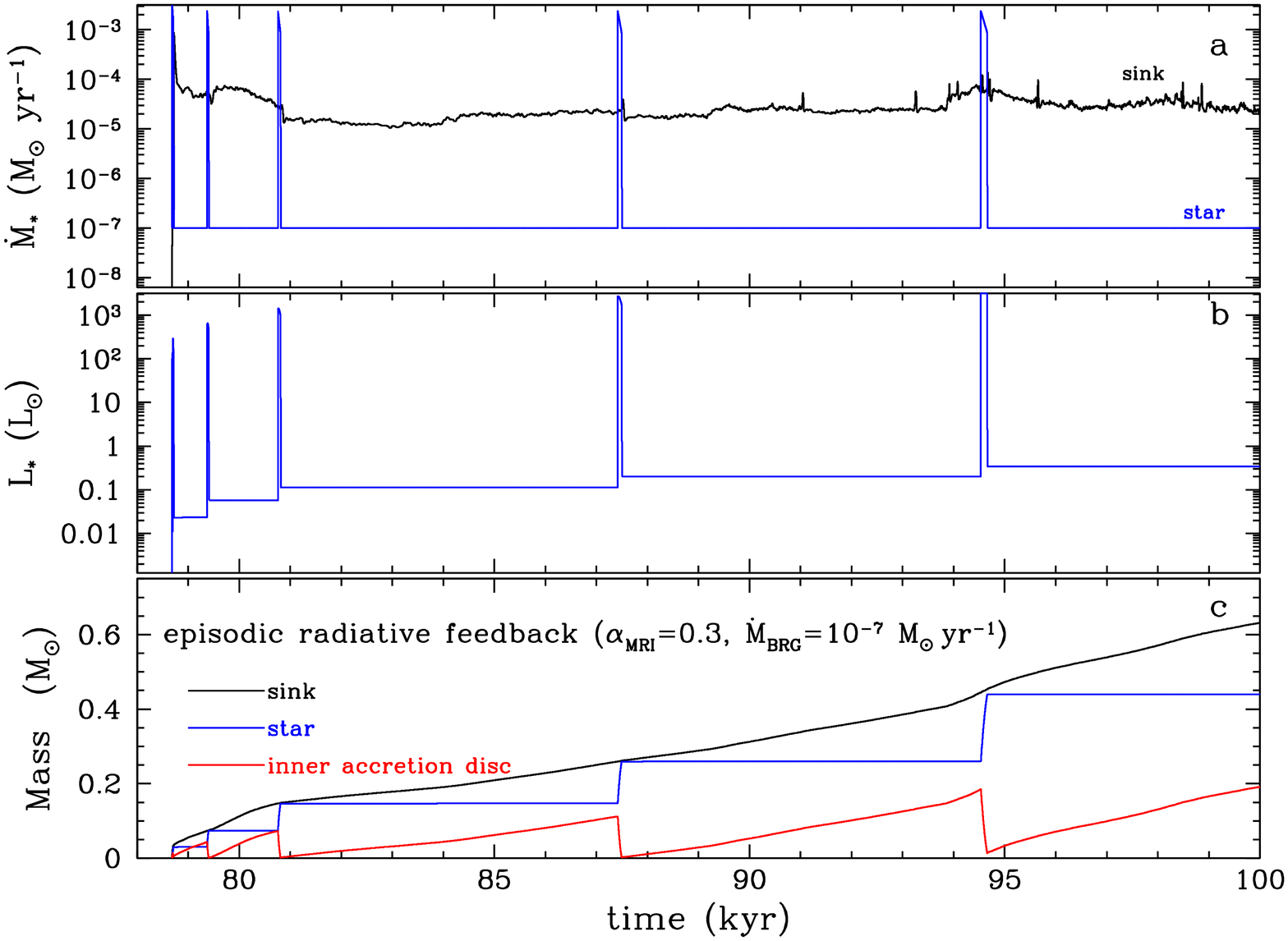}}
\caption{The effect of increasing the viscosity parameter $\alpha_{_{\rm MRI}}$: (a) the accretion rates, (b) the luminosities, and (c) the sink/star/IAD masses of the primary protostar formed in the simulation with episodic accretion  and quiescent accretion of $\dot{M}_{_{\rm BRG}}=10^{-7}\,{\rm M}_{\sun}\,{\rm yr}^{-1}$, (i) with {$\alpha_{_{\rm MRI}}\!=\!0.01$} (ea3; top), (ii)  with {$\alpha_{_{\rm MRI}}\!=\!0.1$} (ea2; middle), and (iii) with {$\alpha_{_{\rm MRI}}\!=\!0.3$} (ea3; bottom). The duration of the episodic accretion events is longer for smaller $\alpha_{_{\rm MRI}}$, and, as a result, the intervals between successive events are shorter, suppressing fragmentation of the disc around the primary protostar.}
\label{fig:group.b}
\end{figure}

\begin{figure}
\centerline{
\includegraphics[height=\columnwidth,angle=-90]{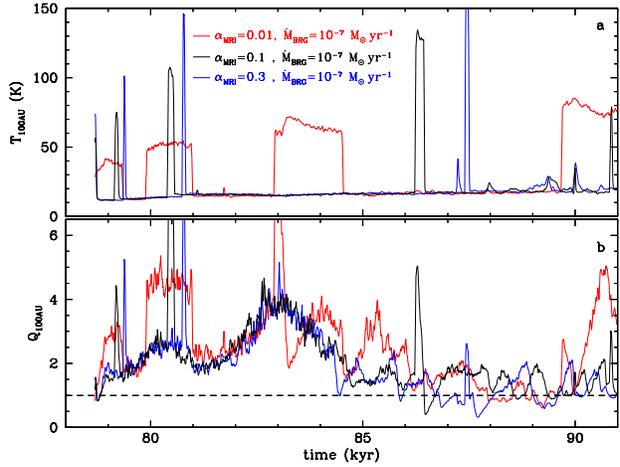}}
\caption{The azimuthally-averaged (a) temperature (top), and (b) Toomre parameter $Q$ (bottom) at radius 100~AU from the primary protostar plotted against time, for runs with different $\alpha_{_{\rm MRI}}$
(red: $\alpha_{_{\rm MRI}}=0.01$, run ea3; black: $\alpha_{_{\rm MRI}}=0.1$, run ea2; blue: $\alpha_{_{\rm MRI}}=0.3$, run ea4). In these runs $\dot{M}_{_{\rm BRG}}=10^{-7}\,{\rm M}_{\sun}\,{\rm yr}^{-1}$.}
\label{fig:toomre.b}
\end{figure}

The viscosity parameter $\alpha_{_{\rm MRI}}$ controls how fast matter accretes onto the protostar when the MRI is active; the higher  $\alpha_{_{\rm MRI}}$ is, the faster angular momentum can be transported outwards in the disc, and the faster matter accretes onto the protostar. Hence, as there is a specific amount of mass that is delivered within each episodic outburst ($M_{\rm _{ MRI}}$; cf.  Eq.~\ref{eq:m.mri}), the outburst duration, $\Delta t_{\rm _{ MRI}}$, is shorter for larger $\alpha_{_{\rm MRI}}$ (cf. Eq.~\ref{eq:mridt}). On the other hand, the outburst is stronger, i.e. the accretion rate onto the the primary protostar and the luminosity during the outburst are larger (Eqs.~\ref{eq:mrimdot},\ref{eq:dmdtdrop}). 

These are evident  in Fig.~\ref{fig:group.b}, where the accretion rate and luminosity of the primary protostar are plotted for 3 runs (ea3, ea2, ea4; top to bottom) with $\alpha_{_{\rm MRI}}\!=\!0.01, 0.1, 0.3$ respectively. In the run with  $\alpha_{_{\rm MRI}}\!=\!0.01$ (ea3) the durations of the episodic accretion events are longer, and disc fragmentation is suppressed. This is because the disc remains hot for longer periods of time, whereas the intervals in between outbursts are rather short and gravitational instabilities cannot fully develop (see Fig.~\ref{fig:toomre.b}). The growth of the disc around the primary protostar is very similar to the case with continuous radiative feedback. 

In  the run with  $\alpha_{_{\rm MRI}}\!=\!0.3$ (ea4) the outbursts are more prominent but much shorter in duration; there is enough time for gravitational instabilities to develop and the disc to fragment. In this case two secondary low-mass H-burning stars form in the disc (see Table~\ref{tab:summary}).

We conclude that the strength of the outbursts is rather irrelevant  regarding disc fragmentation.  The crucial factor that determines whether the disc around the primary protostar fragments is the interval between successive outbursts. Shorter outbursts and longer intervals between them favour disc fragmentation.

\begin{figure}
\centerline{
\includegraphics[height=\columnwidth,angle=-90]{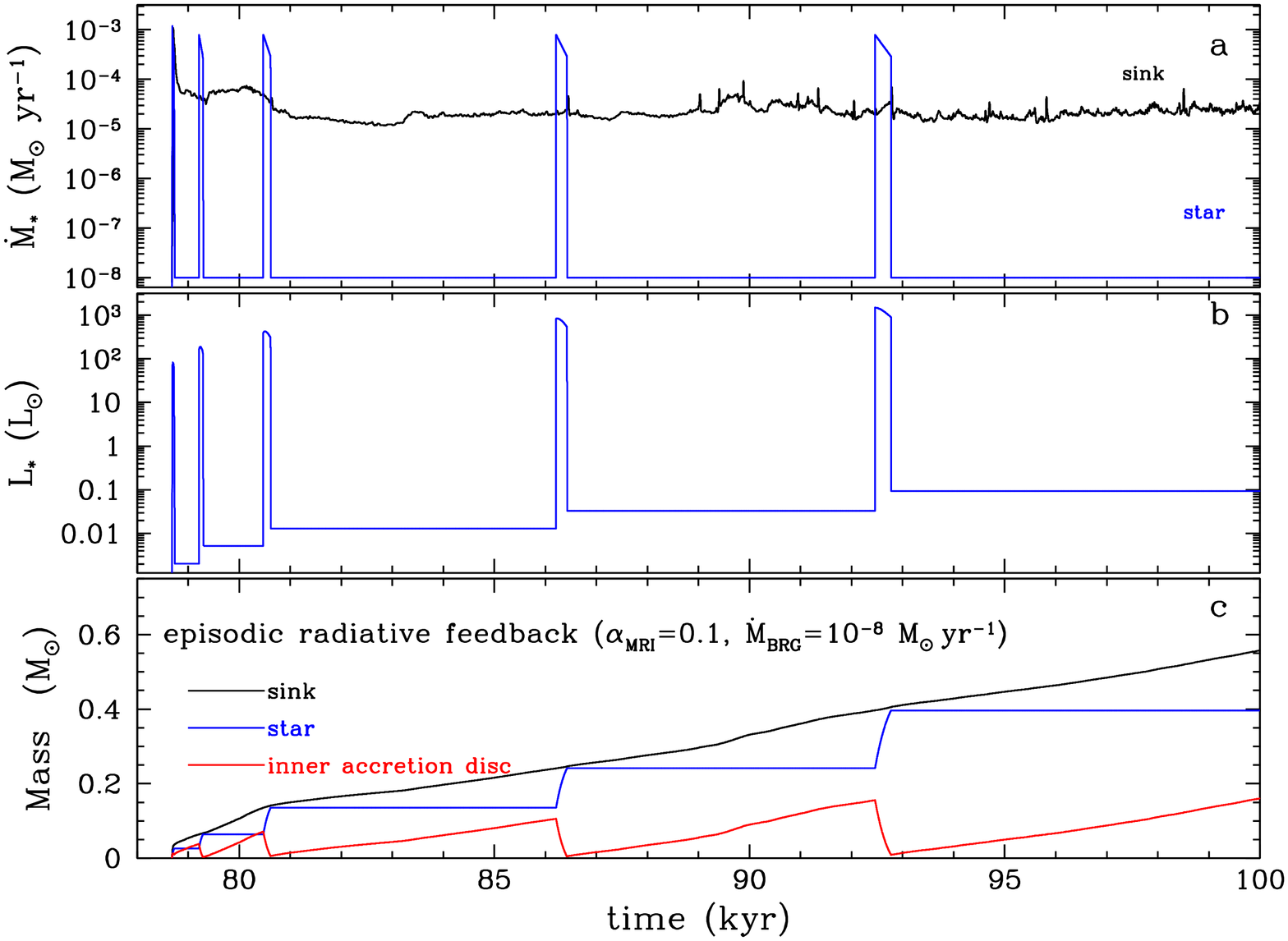}}
\centerline{
\includegraphics[height=\columnwidth,angle=-90]{ea2.eps}}
\centerline{
\includegraphics[height=\columnwidth,angle=-90]{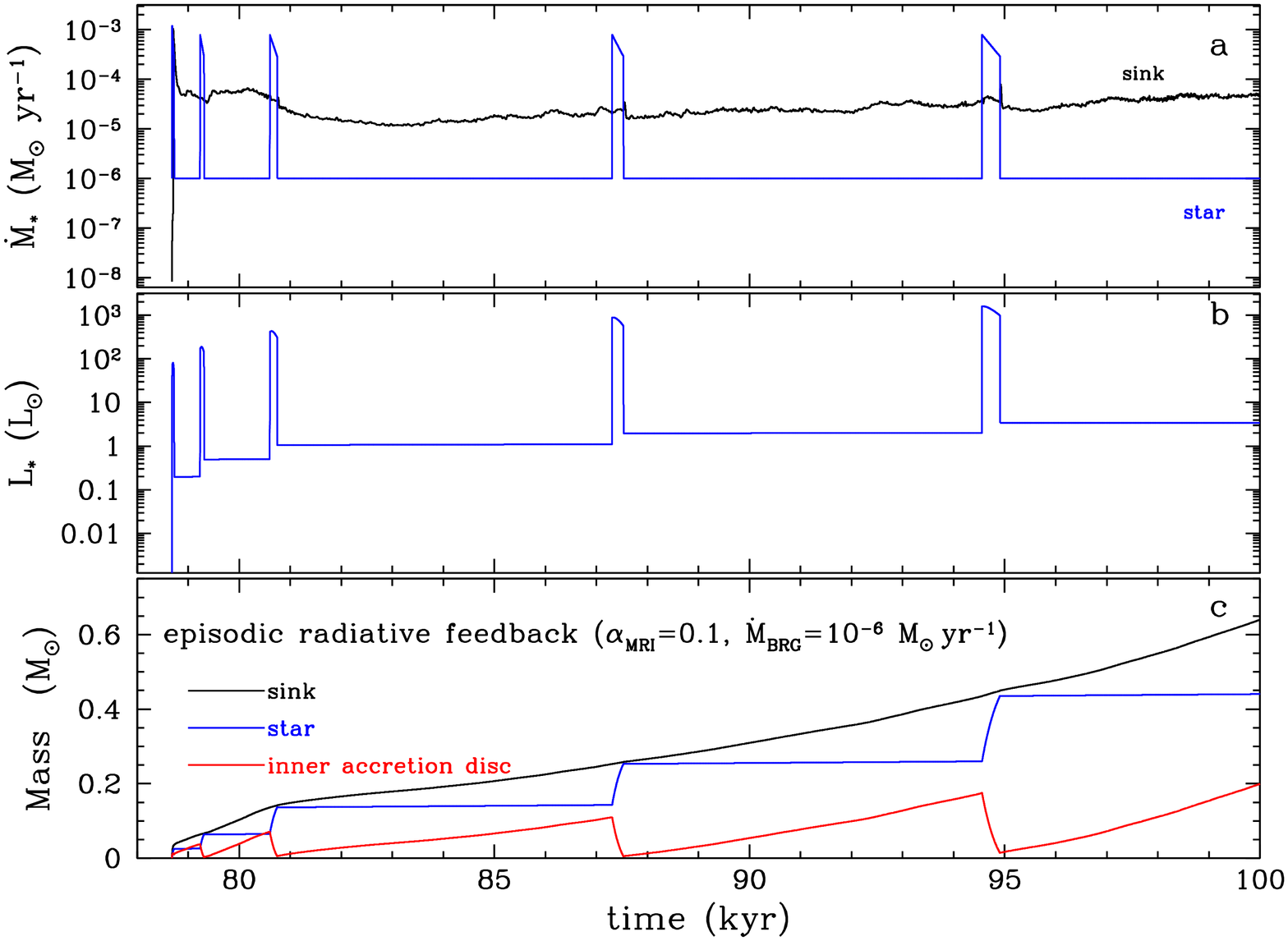}}
\caption{The effect of increasing $\dot{M}_{_{\rm BRG}}$: (a) the accretion rates, (b) the luminosities, and (c) the sink/star/IAD masses of the primary protostars formed in the simulations with episodic accretion  and {$\alpha_{_{\rm MRI}}\!=\!0.1$}, (i) with quiescent accretion rate $\dot{M}_{_{\rm BRG}}=10^{-8}\,{\rm M}_{\sun}\,{\rm yr}^{-1}$ (ea5; top), (ii) with quiescent accretion rate $\dot{M}_{_{\rm BRG}}=10^{-7}\,{\rm M}_{\sun}\,{\rm yr}^{-1}$ (ea2; middle), and (i) with quiescent accretion rate $\dot{M}_{_{\rm BRG}}=10^{-6}\,{\rm M}_{\sun}\,{\rm yr}^{-1}$ (ea6; bottom). The fragmentation of the disc around the primary protostar  is not suppressed in the first two cases, with three and two secondary objects forming, respectively. However, a continuous, quiescent accretion of  $\dot{M}_{_{\rm BRG}}=10^{-6}\,{\rm M}_{\sun}\,{\rm yr}^{-1}$ is enough to raise the luminosity of the primary protostar to $\stackrel{_>}{_\sim} 1~L_{\sun}$  and completely suppress fragmentation.}
\label{fig:group.c}
\end{figure}

\begin{figure}
\centerline{
\includegraphics[height=\columnwidth,angle=-90]{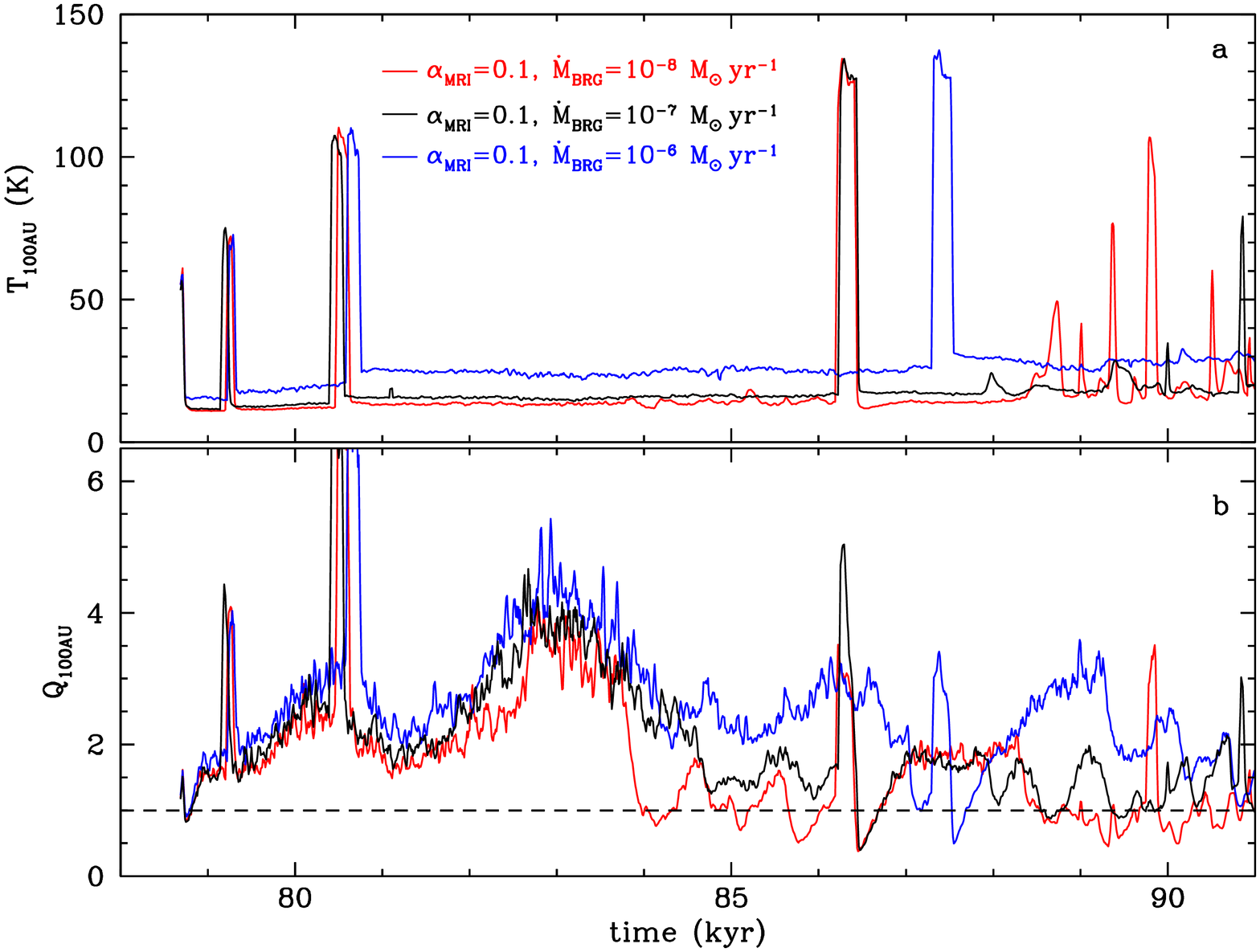}}
\caption{The azimuthally-averaged (a) temperature (top), and (b) Toomre parameter $Q$ (bottom) at radius 100~AU from the primary protostar plotted against time, for runs with different quiescent accretion rates
$\dot{M}_{_{\rm BRG}}$:
(red: $\dot{M}_{_{\rm BRG}}=10^{-8}\,{\rm M}_{\sun}\,{\rm yr}^{-1}$, run ea5; black: $\dot{M}_{_{\rm BRG}}=10^{-6}\,{\rm M}_{\sun}\,{\rm yr}^{-1}$, run ea2; blue: $\dot{M}_{_{\rm BRG}}=10^{-7}\,{\rm M}_{\sun}\,{\rm yr}^{-1}$, run ea6). In  these runs $\alpha_{_{\rm MRI}}\!=\!0.1$. }
\label{fig:toomre.c}
\end{figure}

\subsubsection {$\dot{M}_{_{\rm BRG}}=10^{-8}, 10^{-7},10^{-6}\,{\rm M}_{\sun}\,{\rm yr}^{-1}$
($\alpha_{_{\rm MRI}}\!=\!0.1$) }

The quiescent accretion rate $\dot{M}_{_{\rm BRG}}$ determines how much energy is released by the protostar when the MRI is inactive; the higher the  $\dot{M}_{_{\rm BRG}}$, the higher the luminosity released by the primary protostar, the hotter the attendant  disc becomes. 
 
 This is shown in the luminosity plots for the three runs of increasing   $\dot{M}_{_{\rm BRG}}$ (ea5, ea2, ea6)  in  Fig.~\ref{fig:group.c}. For  $\dot{M}_{_{\rm BRG}}=10^{-8}\,{\rm M}_{\sun}\,{\rm yr}^{-1}$ the luminosity of the protostar when the MRI is inactive is below $0.1~{\rm L}_{\sun}$, whereas in the run with 
 $\dot{M}_{_{\rm BRG}}=10^{-6}\,{\rm M}_{\sun}\,{\rm yr}^{-1}$ the luminosity is $1-2$~${\rm L}_{\sun}$. As a result the disc remains hotter in the second case. In Fig.~\ref{fig:toomre.c} we plot the temperature at 100 AU from the primary protostar; it can be seen that, when the MRI is inactive, in the former run the temperature remains low at about $10-15$~K, whereas in the latter run the temperature is relatively high at $\sim20-30$~K. This temperature is high enough to stabilise the disc and suppress fragmentation.
 
In the run with  $\dot{M}_{_{\rm BRG}}=10^{-8}\,{\rm M}_{\sun}\,{\rm yr}^{-1}$ (ea5) the disc fragmets and two low-mass H-burning stars form, in the run with   $\dot{M}_{_{\rm BRG}}=10^{-7}\,{\rm M}_{\sun}\,{\rm yr}^{-1}$ (ea2) the disc fragments but only one low-mass H-burning stars form, and finally in the run with  $\dot{M}_{_{\rm BRG}}=10^{-6}\,{\rm M}_{\sun}\,{\rm yr}^{-1}$ (ea6) the disc is stabilised by the released energy and it does not fragment.
  
We conclude that high quiescent accretion rate  $\dot{M}_{_{\rm BRG}}$ makes disc fragmentation less favourable, with $\dot{M}_{_{\rm BRG}}\stackrel{>}{_\sim}10^{-6}\,{\rm M}_{\sun}\,{\rm yr}^{-1}$ being enough to totally suppress fragmentation.

\begin{figure}
\centerline{
\includegraphics[width=\columnwidth]{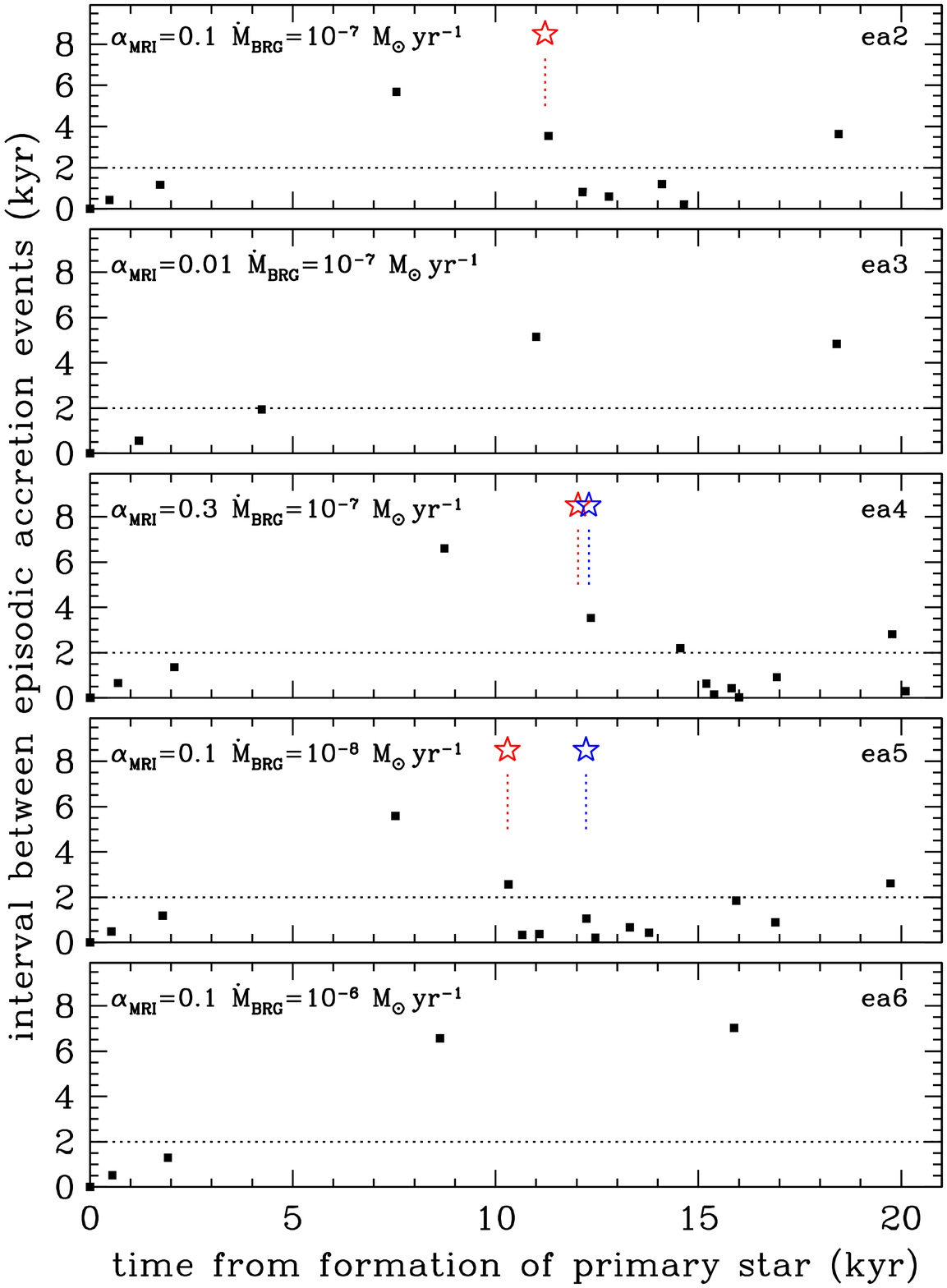}}
\caption{The intervals between successive episodic events (including events from the primary and from the secondary protostars) for runs ea2 to ea6 (top to bottom) plotted against time measured from the formation of the primary protostar. The coloured stars correspond to the times of formation of each one of the secondary protostars, and the horizontal dotted line to an interval of 2 kyr, which is on the order of the dynamical timescale of fragmentation.}
\label{fig:intervals}
\centerline{
\includegraphics[height=\columnwidth,angle=-90]{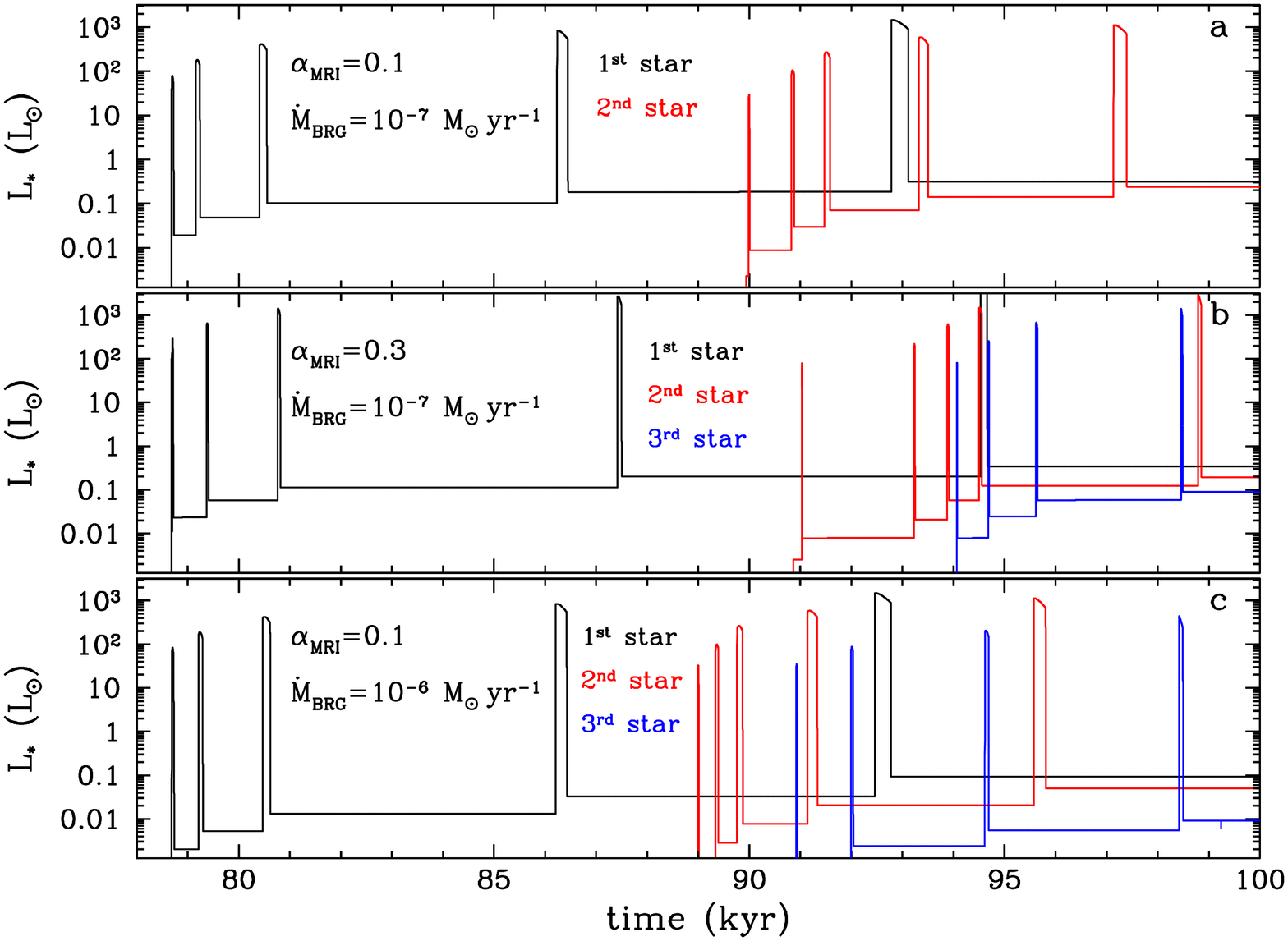}}
\caption{The luminosities of all protostars formed in the simulations where multiple protostars form (top: ea2, middle: ea4, bottom: ea5). After secondary protostars form they also exhibit episodic accretion events and thus they increase the temperature in their immediate vicinity, i.e. the disc of the primary protostar. Therefore after a couple of secondary protostars have formed further fragmentation tends to be suppressed. This is because the intervals between successive luminosity bursts, i.e. when fragmentation is likely, are significantly shortened (cf. Figure~\ref{fig:intervals}).}
\label{fig:secondaries}
\end{figure}

\section{The importance of episodic accretion for disc fragmentation}
\label{sec:discussion}

Radiative feedback from young protostars has the potential to heat and stabilise their discs, suppressing fragmentation \citep{Offner09,Bate09}. Episodic accretion provides a way to limit the effect of protostellar radiative feedback. In this model the accretion of matter onto the protostars occurs not continuously, as previously has been assumed, but in short episodic outbursts. Therefore  energy is released in short batches rather than continuously. 

The crucial factors that determine whether fragmentation is promoted by episodic accretion are (i) the interval between successive outbursts, and (ii) the energy released by the protostar when the outbursts are absent (i.e. when the MRI is inactive). Fragmentation will occur if the disc remains cool enough for  long enough that  GIs have time to grow. Thus, provided that the intervals between successive outbursts are longer than the dynamical timescale (which at $\sim 100$~AU is $\sim1-2$~kyr), and the quiescient accretion rate onto the protostar  $\dot{M}_{_{\rm BRG}}$ is at most a few times $10^{-7}\,{\rm M}_{\sun}\,{\rm yr}^{-1}$, then fragmentation is possible, and indeed we find that it happens (see Table~\ref{tab:summary}). 

This is demonstrated in Fig.~\ref{fig:intervals} where we plot the interval between successive outbursts (taking into account all young protostars in the simulation, i.e. the primary protostar and the secondary protostars formed by disc fragmentation). Long intervals  ($\sim6$~kyr) favour disc fragmentation provided there is still enough mass in the disc; when the intervals become shorter further fragmentation is suppressed (ea2, ea4, ea5). In the last run (ea6) fragmentation does not happen as  $\dot{M}_{_{\rm BRG}}= 10^{-6}\,{\rm M}_{\sun}\,{\rm yr}^{-1}$, and the disc remains relatively hot  ($\sim20-30$~K) and stable, despite the fact that the intervals between successive outbursts are quite long ($\sim6-7$~kyr).

Episodic accretion may also act against fragmentation after a few secondary protostars have formed. These protostars also exhibit episodic outbursts of accretion, leading to outbursts of luminosity, so if they remain in the vicinity of the primary protostar they also heat and stabilise its disc. This is shown in Fig.~\ref{fig:secondaries}, in which we plot the luminosity of all the protostars formed in the simulation (for those simulations where more than one protostar is formed, i.e. ea2, ea4, ea5). We see that after the second protostar has formed the intervals between outbursts become much shorter and fragmentation is therefore suppressed; this effect is further enhanced by the formation of a third protostar, which has its own outbursts, and therefore makes the intervals between outbursts (from {\it all} protostars) even shorter.

We conclude that episodic accretion may initially promote disc fragmentation and the formation of  a couple of secondary low-mass stars (and/or brown dwarfs) in the disc, but suppresses subsequent fragmentation, limiting the number of objects that can form in the disc \citep[cf.][]{Stamatellos09}

\section{Summary}

We have developed a phenomenological model to include the effects of episodic accretion (leading to episodic radiative feedback) in radiative hydrodynamic simulations of star formation. The model is based on the semi-analytical approach of \cite{Zhu10}, and describes the basic aspects of material being episodically accreted onto a protostar and the resulting episodic energy release \citep{Stamatellos11}.

The model has two free parameters, namely the viscosity parameter $\alpha_{_{\rm MRI}}$ which controls how fast matter flows onto the protostar when the MRI is active, and $\dot{M}_{_{\rm BRG}}$ which represents the quiescent accretion rate onto the protostar, when  the MRI is inactive. These parameters are poorly constrained and they may vary from protostar to protostar in different environments. We have shown that if the accretion of material onto protostars is episodic rather than continuous, then radiative feedback from young protostars may not be as effective in suppressing disc fragmentation as previously thought \citep{Offner09,Bate09}, provided (i) that the interval between outbursts is longer than the dynamical timescale for fragmentation, i.e. a few thousand years, and (ii) the quiescent accretion rate onto the protostar is less than a few $10^{-7}\,{\rm M}_{\sun}\,{\rm yr}^{-1}$ .  The time interval between successive outbursts depends on the details of the activation of the MRI, on how fast mass is delivered onto the inner accretion disc, and consequently on how fast mass infalls onto the disc, and on how fast it spirals into the inner disc region (see Eq.~\ref{eq:dtACC}). These physical processes are not well understood.

Adopting typical values for the model parameters we find that disc fragmentation is initially promoted, and a few low-mass secondary objects form; however, further fragmentation is suppressed due to outbursts of energy from the secondary objects formed, limiting the number of objects that form in the disc.

Environmental factors may be important when considering the effect of episodic accretion. For example, in a dense stellar environment perturbations from fly-bys  may strengthen GIs in the disc \citep{Thies10,Forgan10} and deliver more mass to the inner regions triggering episodic accretion events. Therefore it may be expected that in dense stellar environments or in the centres of stellar clusters more outbursts would lead to stable protostellar discs, and consequently  fewer low-mass stars and brown dwarfs being formed. However, for the moment there is no evidence for this \citep{Riaz12}.
 
 We conclude that episodic accretion may provide an effective way to limit the effects of radiative feedback from newly formed protostars on their environments and facilitate disc fragmentation and the formation of low-mass hydrogen-burning stars, brown dwarfs, and planetary-mass objects.

\section*{Acknowledgements}
Simulations were performed using the Cardiff HPC Cluster {\sc Merlin}. Colour plots were produced using {\sc splash} \citep{Price07}. DS and APW acknowledge support by the STFC grant ST/HH001530/1.  DAH is funded by a Leverhulme Trust Research Project Grant (F/00 118/BJ)

\bibliography{bibliography}{}

\begin{thebibliography}{106}
\expandafter\ifx\csname natexlab\endcsname\relax\def\natexlab#1{#1}\fi

\bibitem[{{Alves} {et~al}\mbox{.}(2001){Alves}, {Lada}, \& {Lada}}]{Alves01}
{Alves} J.~F., {Lada} C.~J., {Lada} E.~A., 2001, \nat, 409, 159

\bibitem[{{Andr\'e} {et~al}\mbox{.}(2000){Andr\'e}, {Ward-Thompson}, \&
  {Barsony}}]{Andre00}
{Andr\'e} P., {Ward-Thompson} D., {Barsony} M., 2000, Protostars and Planets
  IV, Eds. Mannings V., Boss A.P., Russell S. S., Univ. Arizona Press, Tuscon,
  AZ, 59

\bibitem[{{Andr\'e} {et~al}\mbox{.}(1996){Andr\'e}, {Ward-Thompson}, \&
  {Motte}}]{Andre96}
{Andr\'e} P., {Ward-Thompson} D., {Motte} F., 1996, \aap, 314, 625

\bibitem[{{Armitage} {et~al}\mbox{.}(2001){Armitage}, {Livio}, \&
  {Pringle}}]{Armitage01}
{Armitage} P.~J., {Livio} M., {Pringle} J.~E., 2001, \mnras, 324, 705

\bibitem[{{Attwood} {et~al}\mbox{.}(2009){Attwood}, {Goodwin}, {Stamatellos},
  \& {Whitworth}}]{Attwood09}
{Attwood} R.~E., {Goodwin} S.~P., {Stamatellos} D., {Whitworth} A.~P., 2009,
  \aap, 495, 201

\bibitem[{{Balbus} \& {Hawley}(1991)}]{Balbus91}
{Balbus} S.~A., {Hawley} J.~F., 1991, \apj, 376, 214

\bibitem[{{Balbus} \& {Hawley}(1998)}]{Balbus98}
{Balbus} S.~A., {Hawley} J.~F., 1998, Reviews of Modern Physics, 70, 1

\bibitem[{{Balsara}(1995)}]{Balsara95}
{Balsara} D.~S., 1995, Journal of Computational Physics, 121, 357

\bibitem[{{Bate}(2009{\natexlab{a}})}]{Bate09b}
{Bate} M.~R., 2009{\natexlab{a}}, \mnras, 392, 590

\bibitem[{{Bate}(2009{\natexlab{b}})}]{Bate09}
{Bate} M.~R., 2009{\natexlab{b}}, \mnras, 392, 1363

\bibitem[{{Bate}(2012)}]{Bate12}
{Bate} M.~R., 2012, \mnras, 419, 3115

\bibitem[{{Bate} {et~al}\mbox{.}(2002){Bate}, {Bonnell}, \& {Bromm}}]{Bate02}
{Bate} M.~R., {Bonnell} I.~A., {Bromm} V., 2002, \mnras, 332, L65

\bibitem[{{Bate} {et~al}\mbox{.}(1995){Bate}, {Bonnell}, \& {Price}}]{Bate95}
{Bate} M.~R., {Bonnell} I.~A., {Price} N.~M., 1995, \mnras, 277, 362

\bibitem[{{Bell} \& {Lin}(1994)}]{Bell94}
{Bell} K.~R., {Lin} D.~N.~C., 1994, \apj, 427, 987

\bibitem[{{Black} \& {Bodenheimer}(1975)}]{Black75}
{Black} D.~C., {Bodenheimer} P., 1975, \apj, 199, 619

\bibitem[{{Boley} {et~al}\mbox{.}(2007){Boley}, {Hartquist}, {Durisen}, \&
  {Michael}}]{Boley07}
{Boley} A.~C., {Hartquist} T.~W., {Durisen} R.~H., {Michael} S., 2007, \apjl,
  656, L89

\bibitem[{{Boley} {et~al}\mbox{.}(2006){Boley}, {Mej{\'{\i}}a}, {Durisen},
  {Cai}, {Pickett}, \& {D'Alessio}}]{Boley06}
{Boley} A.~C., {Mej{\'{\i}}a} A.~C., {Durisen} R.~H., {Cai} K., {Pickett}
  M.~K., {D'Alessio} P., 2006, \apj, 651, 517

\bibitem[{{Bonnell} \& {Bastien}(1992)}]{Bonnell92}
{Bonnell} I., {Bastien} P., 1992, \apjl, 401, L31

\bibitem[{{Boss} \& {Bodenheimer}(1979)}]{Boss79}
{Boss} A.~P., {Bodenheimer} P., 1979, \apj, 234, 289

\bibitem[{{Burkert} \& {Bodenheimer}(2000)}]{Burkert00}
{Burkert} A., {Bodenheimer} P., 2000, \apj, 543, 822

\bibitem[{{Cai} {et~al}\mbox{.}(2008){Cai}, {Durisen}, {Boley}, {Pickett}, \&
  {Mej{\'{\i}}a}}]{Cai08}
{Cai} K., {Durisen} R.~H., {Boley} A.~C., {Pickett} M.~K., {Mej{\'{\i}}a}
  A.~C., 2008, \apj, 673, 1138

\bibitem[{{Calvet} {et~al}\mbox{.}(2004){Calvet}, {Muzerolle}, {Brice{\~n}o},
  {Hern{\'a}ndez}, {Hartmann}, {Saucedo}, \& {Gordon}}]{Calvet04}
{Calvet} N., {Muzerolle} J., {Brice{\~n}o} C., {Hern{\'a}ndez} J., {Hartmann}
  L., {Saucedo} J.~L., {Gordon} K.~D., 2004, \aj, 128, 1294

\bibitem[{{Cartwright} \& {Stamatellos}(2010)}]{Cartwright10}
{Cartwright} A., {Stamatellos} D., 2010, \aap, 516, A99

\bibitem[{{Clarke}(2009)}]{Clarke09}
{Clarke} C.~J., 2009, \mnras, 396, 1066

\bibitem[{{Commer{\c c}on} {et~al}\mbox{.}(2011){Commer{\c c}on}, {Audit},
  {Chabrier}, \& {Chi{\`e}ze}}]{Commercon11}
{Commer{\c c}on} B., {Audit} E., {Chabrier} G., {Chi{\`e}ze} J.-P., 2011, \aap,
  530, A13

\bibitem[{{Dopita}(1978)}]{Dopita78}
{Dopita} A., 1978, \aap, 63, 237

\bibitem[{{Dunham} {et~al}\mbox{.}(2010){Dunham}, {Evans}, {Terebey},
  {Dullemond}, \& {Young}}]{Dunham10}
{Dunham} M.~M., {Evans}, II N.~J., {Terebey} S., {Dullemond} C.~P., {Young}
  C.~H., 2010, \apj, 710, 470

\bibitem[{{Dunham} \& {Vorobyov}(2012)}]{Dunham12}
{Dunham} M.~M., {Vorobyov} E.~I., 2012, \apj, 747, 52

\bibitem[{{Enoch} {et~al}\mbox{.}(2009){Enoch}, {Evans}, {Sargent}, \&
  {Glenn}}]{Enoch09}
{Enoch} M.~L., {Evans}, II N.~J., {Sargent} A.~I., {Glenn} J., 2009, \apj, 692,
  973

\bibitem[{{Evans} {et~al}\mbox{.}(2009){Evans}, {Dunham}, {J{\o}rgensen},
  {Enoch}, {Mer{\'{\i}}n}, {van Dishoeck}, {Alcal{\'a}}, {Myers},
  {Stapelfeldt}, {Huard}, {Allen}, {Harvey}, {van Kempen}, {Blake}, {Koerner},
  {Mundy}, {Padgett}, \& {Sargent}}]{Evans09}
{Evans} N.~J. {et~al.}, 2009, \apjs, 181, 321

\bibitem[{{Forgan} \& {Rice}(2010)}]{Forgan10}
{Forgan} D., {Rice} K., 2010, \mnras, 402, 1349

\bibitem[{{Forgan} {et~al}\mbox{.}(2011){Forgan}, {Rice}, {Cossins}, \&
  {Lodato}}]{Forgan:2011a}
{Forgan} D., {Rice} K., {Cossins} P., {Lodato} G., 2011, \mnras, 410, 994

\bibitem[{{Forgan} {et~al}\mbox{.}(2009){Forgan}, {Rice}, {Stamatellos}, \&
  {Whitworth}}]{Forgan09}
{Forgan} D., {Rice} K., {Stamatellos} D., {Whitworth} A., 2009, \mnras, 394,
  882

\bibitem[{{Goodwin} {et~al}\mbox{.}(2004{\natexlab{a}}){Goodwin}, {Whitworth},
  \& {Ward-Thompson}}]{Goodwin04}
{Goodwin} S.~P., {Whitworth} A.~P., {Ward-Thompson} D., 2004{\natexlab{a}},
  \aap, 414, 633

\bibitem[{{Goodwin} {et~al}\mbox{.}(2004{\natexlab{b}}){Goodwin}, {Whitworth},
  \& {Ward-Thompson}}]{Goodwin04b}
{Goodwin} S.~P., {Whitworth} A.~P., {Ward-Thompson} D., 2004{\natexlab{b}},
  \aap, 423, 169

\bibitem[{{Goodwin} {et~al}\mbox{.}(2006){Goodwin}, {Whitworth}, \&
  {Ward-Thompson}}]{Goodwin06}
{Goodwin} S.~P., {Whitworth} A.~P., {Ward-Thompson} D., 2006, \aap, 452, 487

\bibitem[{{Green} {et~al}\mbox{.}(2011){Green}, {Evans}, {K{\'o}sp{\'a}l}, {van
  Kempen}, {Herczeg}, {Quanz}, {Henning}, {Lee}, {Dunham}, {Meeus}, {Bouwman},
  {van Dishoeck}, {Chen}, {G{\"u}del}, {Skinner}, {Merello}, {Pooley},
  {Rebull}, \& {Guieu}}]{Green11}
{Green} J.~D. {et~al.}, 2011, \apjl, 731, L25

\bibitem[{{Greene} {et~al}\mbox{.}(2008){Greene}, {Aspin}, \&
  {Reipurth}}]{Greene08}
{Greene} T.~P., {Aspin} C., {Reipurth} B., 2008, \aj, 135, 1421

\bibitem[{{Hansen} {et~al}\mbox{.}(2012){Hansen}, {Klein}, {McKee}, \&
  {Fisher}}]{Hansen12}
{Hansen} C.~E., {Klein} R.~I., {McKee} C.~F., {Fisher} R.~T., 2012, \apj, 747,
  22

\bibitem[{{Hartmann} {et~al}\mbox{.}(1997){Hartmann}, {Cassen}, \&
  {Kenyon}}]{Hartmann97}
{Hartmann} L., {Cassen} P., {Kenyon} S.~J., 1997, \apj, 475, 770

\bibitem[{{Hartmann} \& {Kenyon}(1985)}]{Hartman85}
{Hartmann} L., {Kenyon} S.~J., 1985, \apj, 299, 462

\bibitem[{{Hartmann} \& {Kenyon}(1996)}]{Hartmann96}
{Hartmann} L., {Kenyon} S.~J., 1996, \araa, 34, 207

\bibitem[{{Herbig}(1977)}]{Herbig77}
{Herbig} G.~H., 1977, \apj, 217, 693

\bibitem[{{Hubber} {et~al}\mbox{.}(2011{\natexlab{a}}){Hubber}, {Batty},
  {McLeod}, {Whitworth}, {Bisbas}, {Stamatellos}, {Walch}, {Rawiraswattana}, \&
  {Goodwin}}]{Hubber11b}
{Hubber} D. {et~al.}, 2011{\natexlab{a}}, in Astrophysics Source Code Library,
  record ascl:1102.010, p. 2010

\bibitem[{{Hubber} {et~al}\mbox{.}(2011{\natexlab{b}}){Hubber}, {Batty},
  {McLeod}, \& {Whitworth}}]{Hubber11}
{Hubber} D.~A., {Batty} C.~P., {McLeod} A., {Whitworth} A.~P.,
  2011{\natexlab{b}}, \aap, 529, A27

\bibitem[{Isella {et~al}\mbox{.}(2009)Isella, Carpenter, \&
  Sargent}]{Isella:2009a}
Isella A., Carpenter J.~M., Sargent A.~I., 2009, The Astrophysical Journal,
  701, 260

\bibitem[{{Jijina} {et~al}\mbox{.}(1999){Jijina}, {Myers}, \&
  {Adams}}]{Jijina99}
{Jijina} J., {Myers} P.~C., {Adams} F.~C., 1999, \apjs, 125, 161

\bibitem[{{Kenyon} {et~al}\mbox{.}(1990){Kenyon}, {Hartmann}, {Strom}, \&
  {Strom}}]{Kenyon90}
{Kenyon} S.~J., {Hartmann} L.~W., {Strom} K.~M., {Strom} S.~E., 1990, \aj, 99,
  869

\bibitem[{{King} {et~al}\mbox{.}(2007){King}, {Pringle}, \& {Livio}}]{King07}
{King} A.~R., {Pringle} J.~E., {Livio} M., 2007, \mnras, 376, 1740

\bibitem[{{Kirk} {et~al}\mbox{.}(2005){Kirk}, {Ward-Thompson}, \&
  {Andr{\'e}}}]{Kirk05}
{Kirk} J.~M., {Ward-Thompson} D., {Andr{\'e}} P., 2005, \mnras, 360, 1506

\bibitem[{{Kratter} {et~al}\mbox{.}(2008){Kratter}, {Matzner}, \&
  {Krumholz}}]{Kratter08}
{Kratter} K.~M., {Matzner} C.~D., {Krumholz} M.~R., 2008, \apj, 681, 375

\bibitem[{{Krumholz}(2006)}]{Krumholz06}
{Krumholz} M.~R., 2006, \apjl, 641, L45

\bibitem[{{Krumholz}(2011)}]{Krumholz11}
{Krumholz} M.~R., 2011, \apj, 743, 110

\bibitem[{{Krumholz} {et~al}\mbox{.}(2010){Krumholz}, {Cunningham}, {Klein}, \&
  {McKee}}]{Krumholz10}
{Krumholz} M.~R., {Cunningham} A.~J., {Klein} R.~I., {McKee} C.~F., 2010, \apj,
  713, 1120

\bibitem[{{Larson}(1981)}]{Larson81}
{Larson} R.~B., 1981, \mnras, 194, 809

\bibitem[{{Lin} \& {Papaloizou}(1985)}]{Lin85}
{Lin} D.~N.~C., {Papaloizou} J., 1985, in Protostars and planets II (A86-12626
  03-90). Tucson, AZ, University of Arizona Press, 1985, p. 981-1072., {Black}
  D.~C., {Matthews} M.~S., eds., pp. 981--1072

\bibitem[{{Lin} \& {Pringle}(1987)}]{Lin87}
{Lin} D.~N.~C., {Pringle} J.~E., 1987, \mnras, 225, 607

\bibitem[{{Lodato} \& {Rice}(2004)}]{Lodato04}
{Lodato} G., {Rice} W.~K.~M., 2004, \mnras, 351, 630

\bibitem[{{Lodato} \& {Rice}(2005)}]{Lodato05}
{Lodato} G., {Rice} W.~K.~M., 2005, \mnras, 358, 1489

\bibitem[{{Machida} {et~al}\mbox{.}(2010){Machida}, {Inutsuka}, \&
  {Matsumoto}}]{Machida10}
{Machida} M.~N., {Inutsuka} S.-i., {Matsumoto} T., 2010, \apj, 724, 1006

\bibitem[{{Machida} {et~al}\mbox{.}(2011){Machida}, {Inutsuka}, \&
  {Matsumoto}}]{Machida11}
{Machida} M.~N., {Inutsuka} S.-i., {Matsumoto} T., 2011, \apj, 729, 42

\bibitem[{{Masunaga} \& {Inutsuka}(2000)}]{Masunaga00}
{Masunaga} H., {Inutsuka} S., 2000, \apj, 531, 350

\bibitem[{{Matzner} \& {Levin}(2005)}]{Matzner05}
{Matzner} C.~D., {Levin} Y., 2005, \apj, 628, 817

\bibitem[{{Mohanty} {et~al}\mbox{.}(2005){Mohanty}, {Jayawardhana}, \&
  {Basri}}]{Mohanty05}
{Mohanty} S., {Jayawardhana} R., {Basri} G., 2005, \apj, 626, 498

\bibitem[{{Morris} \& {Monaghan}(1997)}]{Morris97}
{Morris} J.~P., {Monaghan} J.~J., 1997, Journal of Computational Physics, 136,
  41

\bibitem[{{Muzerolle} {et~al}\mbox{.}(2005){Muzerolle}, {Luhman},
  {Brice{\~n}o}, {Hartmann}, \& {Calvet}}]{Muzerolle05}
{Muzerolle} J., {Luhman} K.~L., {Brice{\~n}o} C., {Hartmann} L., {Calvet} N.,
  2005, \apj, 625, 906

\bibitem[{{Natta} {et~al}\mbox{.}(2004){Natta}, {Testi}, {Muzerolle},
  {Randich}, {Comer{\'o}n}, \& {Persi}}]{Natta04}
{Natta} A., {Testi} L., {Muzerolle} J., {Randich} S., {Comer{\'o}n} F., {Persi}
  P., 2004, \aap, 424, 603

\bibitem[{{Offner} {et~al}\mbox{.}(2009){Offner}, {Klein}, {McKee}, \&
  {Krumholz}}]{Offner09}
{Offner} S.~S.~R., {Klein} R.~I., {McKee} C.~F., {Krumholz} M.~R., 2009, \apj,
  703, 131

\bibitem[{{Offner} {et~al}\mbox{.}(2010){Offner}, {Kratter}, {Matzner},
  {Krumholz}, \& {Klein}}]{Offner10}
{Offner} S.~S.~R., {Kratter} K.~M., {Matzner} C.~D., {Krumholz} M.~R., {Klein}
  R.~I., 2010, \apj, 725, 1485

\bibitem[{{Offner} \& {McKee}(2011)}]{Offner11}
{Offner} S.~S.~R., {McKee} C.~F., 2011, \apj, 736, 53

\bibitem[{{Palla} \& {Stahler}(1993)}]{Palla93}
{Palla} F., {Stahler} S.~W., 1993, \apj, 418, 414

\bibitem[{{Peneva} {et~al}\mbox{.}(2010){Peneva}, {Semkov}, {Munari}, \&
  {Birkle}}]{Peneva10}
{Peneva} S.~P., {Semkov} E.~H., {Munari} U., {Birkle} K., 2010, \aap, 515, A24

\bibitem[{{Price}(2007)}]{Price07}
{Price} D.~J., 2007, \pasa, 24, 159

\bibitem[{{Pudritz} {et~al}\mbox{.}(2007){Pudritz}, {Ouyed}, {Fendt}, \&
  {Brandenburg}}]{Pudritz07}
{Pudritz} R.~E., {Ouyed} R., {Fendt} C., {Brandenburg} A., 2007, Protostars and
  Planets V, Eds. Reipurth B., Jewitt D., Keil K., Univ. Arizona Press, Tucson,
  277

\bibitem[{{Reipurth}(1989)}]{Reipurth89}
{Reipurth} B., 1989, \nat, 340, 42

\bibitem[{{Riaz} {et~al}\mbox{.}(2012){Riaz}, {Lodieu}, {Goodwin},
  {Stamatellos}, \& {Thompson}}]{Riaz12}
{Riaz} B., {Lodieu} N., {Goodwin} S., {Stamatellos} D., {Thompson} M., 2012,
  \mnras, 420, 2497

\bibitem[{{Rice} \& {Armitage}(2009)}]{Rice09}
{Rice} W.~K.~M., {Armitage} P.~J., 2009, \mnras, 396, 2228

\bibitem[{{Shakura} \& {Sunyaev}(1973)}]{Shakura73}
{Shakura} N.~I., {Sunyaev} R.~A., 1973, \aap, 24, 337

\bibitem[{{Shang} {et~al}\mbox{.}(2007){Shang}, {Li}, \& {Hirano}}]{Shang07}
{Shang} H., {Li} Z.-Y., {Hirano} N., 2007, Protostars and Planets V., Eds.
  Reipurth B., Jewitt D., Keil K., Univ. Arizona Press, Tucson, 261

\bibitem[{{Stamatellos} {et~al}\mbox{.}(2007{\natexlab{a}}){Stamatellos},
  {Hubber}, \& {Whitworth}}]{Stamatellos07b}
{Stamatellos} D., {Hubber} D.~A., {Whitworth} A.~P., 2007{\natexlab{a}},
  \mnras, 382, L30

\bibitem[{{Stamatellos} \& {Whitworth}(2003)}]{Stamatellos03}
{Stamatellos} D., {Whitworth} A.~P., 2003, \aap, 407, 941

\bibitem[{{Stamatellos} \& {Whitworth}(2008)}]{Stamatellos08}
{Stamatellos} D., {Whitworth} A.~P., 2008, \aap, 480, 879

\bibitem[{{Stamatellos} \& {Whitworth}(2009{\natexlab{a}})}]{Stamatellos09}
{Stamatellos} D., {Whitworth} A.~P., 2009{\natexlab{a}}, \mnras, 392, 413

\bibitem[{{Stamatellos} \& {Whitworth}(2009{\natexlab{b}})}]{Stamatellos09b}
{Stamatellos} D., {Whitworth} A.~P., 2009{\natexlab{b}}, \mnras, 400, 1563

\bibitem[{{Stamatellos} {et~al}\mbox{.}(2004){Stamatellos}, {Whitworth},
  {Andr{\'e}}, \& {Ward-Thompson}}]{Stamatellos04}
{Stamatellos} D., {Whitworth} A.~P., {Andr{\'e}} P., {Ward-Thompson} D., 2004,
  \aap, 420, 1009

\bibitem[{{Stamatellos} {et~al}\mbox{.}(2007{\natexlab{b}}){Stamatellos},
  {Whitworth}, {Bisbas}, \& {Goodwin}}]{Stamatellos07}
{Stamatellos} D., {Whitworth} A.~P., {Bisbas} T., {Goodwin} S.,
  2007{\natexlab{b}}, \aap, 475, 37

\bibitem[{{Stamatellos} {et~al}\mbox{.}(2011){Stamatellos}, {Whitworth}, \&
  {Hubber}}]{Stamatellos11}
{Stamatellos} D., {Whitworth} A.~P., {Hubber} D.~A., 2011, \apj, 730, 32

\bibitem[{{Tan} \& {McKee}(2004)}]{Tan04}
{Tan} J.~C., {McKee} C.~F., 2004, \apj, 603, 383

\bibitem[{{Terebey} {et~al}\mbox{.}(1984){Terebey}, {Shu}, \&
  {Cassen}}]{Terebey84}
{Terebey} S., {Shu} F.~H., {Cassen} P., 1984, \apj, 286, 529

\bibitem[{{Thies} {et~al}\mbox{.}(2010){Thies}, {Kroupa}, {Goodwin},
  {Stamatellos}, \& {Whitworth}}]{Thies10}
{Thies} I., {Kroupa} P., {Goodwin} S.~P., {Stamatellos} D., {Whitworth} A.~P.,
  2010, \apj, 717, 577

\bibitem[{{Urban} {et~al}\mbox{.}(2010){Urban}, {Martel}, \& {Evans}}]{Urban10}
{Urban} A., {Martel} H., {Evans} N.~J., 2010, \apj, 710, 1343

\bibitem[{{Vorobyov} \& {Basu}(2005)}]{Vorobyov05}
{Vorobyov} E.~I., {Basu} S., 2005, \apjl, 633, L137

\bibitem[{{Walch} {et~al}\mbox{.}(2012){Walch}, {Whitworth}, \&
  {Girichidis}}]{Walch12}
{Walch} S., {Whitworth} A.~P., {Girichidis} P., 2012, \mnras, 419, 760

\bibitem[{{Ward-Thompson} {et~al}\mbox{.}(1999){Ward-Thompson}, {Motte}, \&
  {Andr\'e}}]{WardThompson99}
{Ward-Thompson} D., {Motte} F., {Andr\'e} P., 1999, \mnras, 305, 143

\bibitem[{{Ward-Thompson} {et~al}\mbox{.}(1994){Ward-Thompson}, {Scott},
  {Hills}, \& {Andr\'e}}]{WardThompson94}
{Ward-Thompson} D., {Scott} P.~F., {Hills} R.~E., {Andr\'e} P., 1994, \mnras,
  268, 276

\bibitem[{{Whitehouse} \& {Bate}(2006)}]{Whitehouse06}
{Whitehouse} S.~C., {Bate} M.~R., 2006, \mnras, 367, 32

\bibitem[{{Whitworth} {et~al}\mbox{.}(2007){Whitworth}, {Bate}, {Nordlund},
  {Reipurth}, \& {Zinnecker}}]{Whitworth07}
{Whitworth} A., {Bate} M.~R., {Nordlund} {\AA}., {Reipurth} B., {Zinnecker} H.,
  2007, Protostars and Planets V., Eds. Reipurth B., Jewitt D., Keil K., Univ.
  Arizona Press, Tucson, 459

\bibitem[{{Whitworth} \& {Stamatellos}(2006)}]{Whitworth06}
{Whitworth} A.~P., {Stamatellos} D., 2006, \aap, 458, 817

\bibitem[{{Wilkins} \& {Clarke}(2012)}]{Wilkins12}
{Wilkins} D.~R., {Clarke} C.~J., 2012, \mnras, 419, 3368

\bibitem[{{Young} {et~al}\mbox{.}(2012){Young}, {Bertram}, {Moeckel}, \&
  {Clarke}}]{Young:2012a}
{Young} M.~D., {Bertram} E., {Moeckel} N., {Clarke} C.~J., 2012, ArXiv e-prints

\bibitem[{{Zhu} {et~al}\mbox{.}(2007){Zhu}, {Hartmann}, {Calvet}, {Hernandez},
  {Muzerolle}, \& {Tannirkulam}}]{Zhu07}
{Zhu} Z., {Hartmann} L., {Calvet} N., {Hernandez} J., {Muzerolle} J.,
  {Tannirkulam} A., 2007, \apj, 669, 483

\bibitem[{{Zhu} {et~al}\mbox{.}(2009{\natexlab{a}}){Zhu}, {Hartmann}, \&
  {Gammie}}]{Zhu09}
{Zhu} Z., {Hartmann} L., {Gammie} C., 2009{\natexlab{a}}, \apj, 694, 1045

\bibitem[{{Zhu} {et~al}\mbox{.}(2010{\natexlab{a}}){Zhu}, {Hartmann}, \&
  {Gammie}}]{Zhu10b}
{Zhu} Z., {Hartmann} L., {Gammie} C., 2010{\natexlab{a}}, \apj, 713, 1143

\bibitem[{{Zhu} {et~al}\mbox{.}(2009{\natexlab{b}}){Zhu}, {Hartmann}, {Gammie},
  \& {McKinney}}]{Zhu09b}
{Zhu} Z., {Hartmann} L., {Gammie} C., {McKinney} J.~C., 2009{\natexlab{b}},
  \apj, 701, 620

\bibitem[{{Zhu} {et~al}\mbox{.}(2010{\natexlab{b}}){Zhu}, {Hartmann}, {Gammie},
  {Book}, {Simon}, \& {Engelhard}}]{Zhu10}
{Zhu} Z., {Hartmann} L., {Gammie} C.~F., {Book} L.~G., {Simon} J.~B.,
  {Engelhard} E., 2010{\natexlab{b}}, \apj, 713, 1134

\bibitem[{{Zhu} {et~al}\mbox{.}(2012){Zhu}, {Hartmann}, {Nelson}, \&
  {Gammie}}]{Zhu12}
{Zhu} Z., {Hartmann} L., {Nelson} R.~P., {Gammie} C.~F., 2012, \apj, 746, 110

\end{thebibliography}

\label{lastpage}
\end{document}